# Uplink User-Assisted Relaying in Cellular Networks

Hussain Elkotby, *Student Member IEEE* and Mai Vu, *Senior Member IEEE*

***Abstract*—We use stochastic geometry to analyze the performance of a partial decode-and-forward (PDF) relaying scheme applied in a user-assisted relaying setting, where an active user relays data through another idle user in uplink cellular communication. We present the geometric model of a network deploying user-assisted relaying and propose two geometric cooperation policies for fast and slow fading channels. We analytically derive the cooperation probability for both policies. This cooperation probability is further used in the analytical derivation of the moments of inter-cell interference power caused by system-wide deployment of this user-assisted PDF relaying. We then model the inter-cell interference power statistics using the Gamma distribution by matching the first two moments analytically derived. This cooperation and interference analysis provides the theoretical basis for quantitatively evaluating the performance impact of user-assisted relaying in cellular networks. We then numerically evaluate the average transmission rate performance and show that user-assisted relaying can significantly improve per-user transmission rate despite of increased inter-cell interference. This transmission rate gain is significant for active users near the cell edge and further increases with higher idle user density, supporting user-assisted relaying as a viable solution to crowded population areas.**

***Keywords: user-assisted relaying; partial decode-and-forward; stochastic geometry.***

## I. Introduction

MOBILE operators driven by the increasing number of subscribers and continual customer demand for new and better services place pressing requirements on the underlying wireless technologies to provide high data rates and wide coverage. Future generation networks that promise higher data rates and multifold increase in system capacity include 3GPP Long Term Evolution−Advanced (LTE-A, 4G) and the emerging 5G systems. The fourth generation (4G) wireless systems were designed to fulfill the requirements of the International Mobile Telecommunications − Advanced (IMT-A). LTE as a practical 4G wireless system has been recently deployed in some countries and LTE-A is expected to be deployed soon around the globe [1]. It is well established that 4G networks have just reached the theoretical limit on the data rate with current technologies. These technologies are being complemented in the fifth generation (5G) wireless systems by designing and developing new radio concepts to accommodate higher data rates, larger network capacity, higher energy efficiency, and higher mobility necessary to meet the new and challenging requirements of new wireless applications. 5G wireless systems are expected to support peak data rate of 10 Gb/s for low mobility and 1 Gb/s for high mobility. These networks are expected to be standardized and deployed around and beyond 2020. Various promising technologies are proposed for 5G wireless communication systems such as massive MIMO, energy-efficient communications, Device-to-Device (D2D) communications, millimeter-wave (mmWave), and cognitive radio networks [1]–[3].

### A. Background and Related Works

D2D and Relaying cooperative communications will play important roles in future generations wireless networks. D2D communications enable two proximity users to transmit signal directly without going through the base station; subsequently, 5G wireless systems are expected to relax the restrictions on the need to route all user data through the core network. D2D communications can increase network spectrum utilization and energy efficiency, reduce transmission delay, offload traffic for the base station, and alleviate congestion in the cellular core networks, which make it a promising technology for future wireless systems [3]–[5]. Relay-aided cooperative communication techniques represent another promising technology that improves performance in poor coverage areas by enabling ubiquitous coverage even for users in the most unfavorable channel conditions. The latest release of the LTE standard allows the deployment of fixed wireless relays to help cell-edge mobiles. Yet, other advanced cellular relaying modes are expected in 5G systems to improve the topology and robustness of a cellular network and decrease power consumption. These new technologies include mobile relaying, multi-hop relaying, and user-equipment based (user-assisted) relaying enabled by D2D communications [4]–[7].

Several modes of relay-aided communication have been studied in the literature, including fixed relay station, mobile relay station, and using other user equipment (UE) as relay nodes [8]–[10]. Most existing results are derived for the first two modes of fixed and mobile relay stations. For example, simulation is used to compare the difference between relaying network architectures with mobile or fixed relay stations and contrast their performance gains in [8]. Resource allocation for uplink OFDMA-based cooperative relay networks is studied in [9]. The third mode of relaying through other idle UEs in a cellular network (or user-assisted relaying) has only been studied through system simulations for decode-and-forward relaying in [10].

User-assisted relaying, nevertheless, provides more flexibility than fixed relaying in expanding the base station (BS) coverage into obscured areas, especially where there is high density of idle UEs [7]. The density of idle UEs that are willing to participate in relaying transmission is expected to increase over time with the development of novel pricing models to tempt devices to participate in this type of cooperation [5]. The issue of battery power drainage of mobile UEs due to their

 The authors are with the Department of Electrical and Computer Engineering, Tufts University, Medford, MA, USA. Emails: Hussain.Elkotby@Tufts.edu, Mai.Vu@Tufts.edu

cooperation in relaying other users data to the base station has also been examined through the emerging energy harvesting techniques [11]. In this paper, we analyze the performance of user-assisted relaying when deployed system-wide in a cellular network.

For cellular network analysis, stochastic geometry has been shown to be analytically tractable and able to capture some of the main performance trends. Stochastic geometry is used to develop a tractable model for downlink heterogeneous cellular networks which is comparable to the grid model and actual 4G deployment data in [12]. This model is further used to analyze downlink coordinated multipoint beamforming, in which each user equipped with a single antenna can be served by either one or two base stations connected over backhaul links of infinite capacity [13]. A user decides whether to connect to one or two base stations based on geometric policies taking into account its relative distances to the two closest base stations. Poisson spatial distribution is used to develop an analytic interference model for multi-cell multiple-input multiple-output cellular networks and derive its downlink average capacity [14]. Stochastic geometry is also used to analyze the performance of decode-and-forward relaying techniques in uplink cellular networks, under the specific setting of a fixed number of relays deployed at a fixed distance from the BS with equal angular separation in each cell [15]. Recently, stochastic geometry is used to analyze the outage performance of a large decentralized wireless network in which transmitters may be aided by nearby relays using a full-duplex decode-and-forward scheme [16]. A transmitter chooses the relaying node as the nearest neighbor within a cone with an aperture angle centered towards the destination. The outage performance is then used to design an on/off relay activation strategy that depends on parameters such as: path loss exponent, source-destination distance, cone aperture angle, transmitting nodes density, and the attempted transmission rate.

### B. Main Results and Contributions

In this paper, we study the performance of a partial decode-and-forward (PDF) user-assisted relaying scheme in uplink cellular networks. To the best of our knowledge, our work is the first that analyzes user-assisted relaying in a network-wide cellular context. The main question under consideration is how network-wide deployment of user-assisted relaying affects the system performance. Since some idle users are now transmitting by relaying information of other users, the amount of interference generated to the network will increase. We use stochastic geometry as a tool to model and analyze this interference as well as the cooperation policy which governs how to select the idle user to act as a relay. In this work, we consider the practical strategy where each active UE selects the closest idle UE to relay its message to the destination, given that D2D communications will be enabled between UEs that are in proximity of each other. Other more complex policies such as selecting the idle user with the strongest link instead of the closest one or the idle user resulting in the highest relayed data rate, can also be considered in our proposed framework and are left as future work. As a base for analysis, we assume all nodes are equipped with a single antenna. The multiple antenna relay channel is still a topic of current research, where the optimal decode-and-forward input for a stand-alone MIMO relay channel is only recently investigated in [17]. Our work provides the basis for later extension of network-wide deployment to the multiple antenna case.

We provide the geometric basis for a rigorous analysis of performance metrics such as the average throughput for the whole cellular system in contrast to the stand-alone analysis in [18]. The contributions and novelties of this paper are summarized as follows:

1) We propose a geometric model for user-assisted relaying in uplink cellular networks. The model assumes a reuse factor of one in the whole network. We consider scenarios where there are multiple idle users as potential candidates for helping an active user per resource block in each cell, indicated by the ratio between idle and active user densities.
2) We propose two practical cooperation policies as compared to the ideal policy presented in [18]: a pure geometric policy for fast fading channels and a hybrid fading and geometric policy for slow fading channels. Further, we analytically derive the cooperation probability for these policies and compare them with simulation results of the ideal case when the nodes know the channel state information perfectly, and show that these policies have similar cooperation probabilities.
3) We analytically formulate the out-of-cell interference power at both the destination base station and the relaying user within a cell of a given radius. This formulation takes into account the random locations of all users and base stations, as well as both large scale fading due to path loss and small scale fading due to the constructive and destructive sum of the multiple signal paths between a transmitter and a receiver. We then derive the Laplace transforms of these interference powers, which allow us to analytically compute any moments of the interference. We further specifically compute the first and second moments of this out-of-cell interference power in closed-form.
4) We use second moment matching to model the out-of-cell interference power as a Gamma distribution and numerically evaluate the fit of this model. Results show that the Gamma distribution provides a good fit within the range of regulated transmit power and for all range of user-base station distances. This result therefore provides a tractable analytic model for the out-of-cell interference power generated by network-wide deployment of user-assisted relaying.
5) We use the developed analytical model of interference power to numerically evaluate system performance and provide a quantitative analysis of uplink user-assisted relaying using the average per-user data rate as the metric. Results show that the rate gain is significant when the active user is located in the one-half or one-third ring near the cell edge. The average rate gain increases with higher idle user density and can be up to 50% when idle users are six times denser than active users. The maximum gain can be as high as 200% when the idle user is ideally located about half way between the active user and the base station.

The remainder of this paper is organized as follows: Section II describes the PDF relaying transmission scheme. Section III introduces the network geometric and interference models. Section IV introduces the cooperation policies and analyzes their probabilities. Section V provides the interference power analysis. In Section VI, we evaluate our geometric model and discuss the validity of interference Gamma distribution approximation. Section VII shows the numerical results of system performance. Finally, Section VIII presents our conclusion.

## II. Partial Decode-and-Forward Relaying Scheme

In this section, we describe the stand-alone half-duplex PDF relaying scheme as proposed in [18]. We discuss the signal design, decoding techniques and achievable rate of this relaying scheme. We then formulate the achievable rate when the scheme is deployed according to a deployment policy. The discussion here provides the basis for subsequent network deployment.

### A. Standard Channel Model

In Fig. 1, we consider the stand-alone half-duplex relay channel consisting of a single set of source $\mathcal{S}$, relay $\mathcal{R}$ and destination $\mathcal{D}$. In this model, each transmission block is divided into 2 phases where we assume flat fading over the two phase period. We model the received signals at $\mathcal{R}$ and $\mathcal{D}$, respectively, during the first phase as

$$Y_r^b = h_{sr} x_s^b + Z_r^b, \qquad Y_d^b = h_{sd} x_s^b + Z_d^b \tag{1}$$

where $b$ stands for broadcast transmission in which $\mathcal{S}$ broadcasts to both $\mathcal{R}$ and $\mathcal{D}$. The signal $x_s^b$ is the transmitted codeword from $\mathcal{S}$ in the first phase; $Z_r^b$ and $Z_d^b$ are $i.i.d$ $\mathcal{CN}(0, \sigma^2)$ that represent the noise at $\mathcal{R}$ and $\mathcal{D}$; and $h_{sr}$ and $h_{sd}$ are the $\mathcal{S}$-to-$\mathcal{R}$ and $\mathcal{S}$-to-$\mathcal{D}$ channels, respectively.

Also, we can model the received signal at $\mathcal{D}$ during the second phase as

$$Y_d^m = h_{sd} x_s^m + h_{rd} x_r^m + Z_d^m \tag{2}$$

where $m$ denotes multiple access transmission in which both $\mathcal{S}$ and $\mathcal{R}$ send information to $\mathcal{D}$. The signals $x_s^m$ and $x_r^m$ are the transmitted codewords from $\mathcal{S}$ and $\mathcal{R}$ in the second phase; $h_{rd}$ is the $\mathcal{R}$-to-$\mathcal{D}$ channel; and $Z_d^m \sim \mathcal{CN}(0, \sigma^2)$ represents the noise at $\mathcal{D}$.

Similarly, we model the received signal at $\mathcal{D}$ in the direct transmission scheme as

$$Y_d = h_{sd} x_s + Z_d \tag{3}$$

where $x_s$ is the transmitted codeword from $\mathcal{S}$ in direct transmission; and $Z_d \sim \mathcal{CN}(0, \sigma^2)$ represents the noise at $\mathcal{D}$ in the direct transmission case.

All the channels $h_{xy}, xy \in \{sr, sd, rd\}$ are complex channel gains with uniformly distributed phases that capture both the small and large scale fading and can be written in the form $h_{xy} = e^{j\theta_l} |h_{xy}|$ where $\theta_l \sim \mathcal{U}[0, 2\pi], l \in \{1, 2, r\}$, respectively.

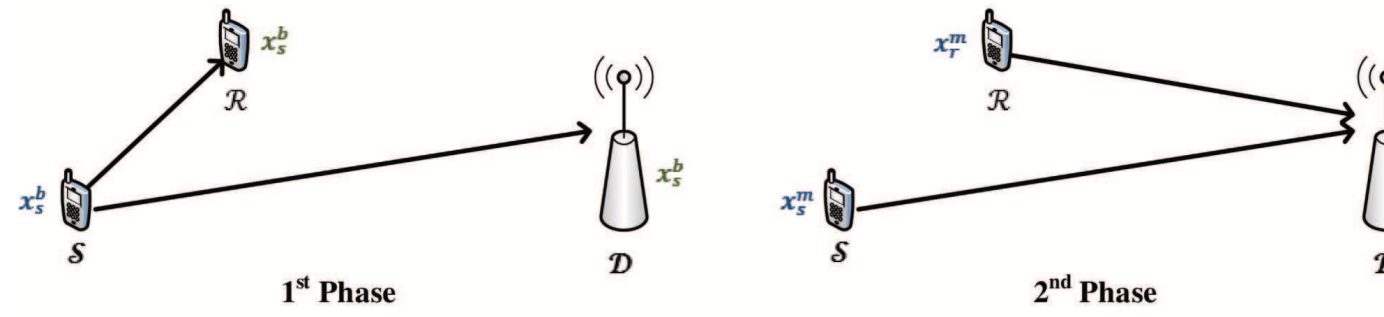

Fig. 1: Relay scheme and channel model

### B. Coherent PDF Relaying Scheme

In each transmission block, $\mathcal{S}$ uses superposition coding and splits its information into a common part at rate $R_1$ and a private part at rate $R_2$. The common part is encoded via $U_s^b$ in the $1^{st}$ phase and $U_s^{m1}$ in the $2^{nd}$ phase; and the private part is encoded via $V_s^{m2}$ in the $2^{nd}$ phase. The relay $\mathcal{R}$ encodes the same common information using $U_s^{m1}$ in the $2^{nd}$ phase.

Specifically, using Gaussian signaling, $\mathcal{S}$ and $\mathcal{R}$ constructs their transmit signals as follows:

$$\text{Phase 1:} \quad x_s^b = \sqrt{P_s^b} U_s^b, \tag{4}$$

$$\begin{aligned} \text{Phase 2:} \quad x_r^m &= \sqrt{P_r^m} U_s^{m1}, \\ x_s^m &= \sqrt{P_s^{m1}} U_s^{m1} + \sqrt{P_s^{m2}} V_s^{m2} \end{aligned} \tag{5}$$

where the codewords $U_s^b$, $U_s^{m1}$, and $V_s^{m2}$ are independent standard Gaussian with zero mean and unit variances. Note that codewords $U_s^b$ and $U_s^{m1}$ both encode the same common information but are generated independently of each other; that is, they are picked from independent codebooks or constellations. The transmit power of $\mathcal{S}$ and $\mathcal{R}$ need to satisfy the average power constraints:

$$\alpha_1 P_s^b + \alpha_2 P_s^m = P_s, \quad \alpha_2 P_r^m = P_r, \tag{6}$$

where $P_s$ and $P_r$ represent the total power allocated to the source and relay nodes within a single transmission period; $\alpha_1$ and $\alpha_2 = 1 - \alpha_1$ represent the portions of the transmission time allocated to the first and second phases, respectively.

The relay $\mathcal{R}$ decodes $U_s^b$ at the end of the $1^{st}$ phase and encodes the just decoded common information using $U_s^{m1}$ in the $2^{nd}$ phase. At the end of the $2^{nd}$ phase, $\mathcal{D}$ utilizes both received signals, $Y_d^b$ in eq. (1) and $Y_d^m$ in eq. (2), to decode both the common and private parts using joint maximum likelihood (ML) decoding rule over both phases.

Knowledge about the phase offset between the two transmitting nodes - the source and the relay - to the destination is assumed in the second phase of this transmission scheme in order to achieve coherent source-relay transmission, as usually done in the literature [19]–[21]. This assumption can be further justified in our model by noting that phase offset between the two transmitting nodes can be estimated at the destination base station and fed-back using a dedicated resource. Another more appealing technique is to benefit from the channel reciprocity to estimate the channels at the transmitting nodes and exploit the fact that D2D communications in future cellular networks are expected between two devices separated by short distances; therefore reliable communication links can be established over some dedicated resource blocks at low transmit powers in order to send the required transmit phase information. Such phase information exchange allows the extensive reuse of the resources among different D2D connections within the same cell because of the proximity of such exchanges [22].

### C. Transmission Schemes Achievable Rates

With transmit signals in Eqs. (4)−(5) and joint ML decoding rule at $\mathcal{D}$, we obtain the following achievable rate for PDF relaying which ensures reliable decoding at $\mathcal{R}$ and $\mathcal{D}$:

$$R_{PDF} \leq \min(C_1 + C_2, C_3), \tag{7}$$

where
$$\begin{aligned} C_1 &= \alpha_1 \log\left(1 + |h_{sr}|^2 P_s^b\right), \\ C_2 &= \alpha_2 \log\left(1 + |h_{sd}|^2 P_s^{m2}\right), \\ C_3 &= \alpha_1 \log\left(1 + |h_{sd}|^2 P_s^b\right) + \alpha_2 \log\bigg(1 + |h_{sd}|^2 P_s^{m2} \\ &\quad + \left(|h_{sd}|\sqrt{P_s^{m1}} + |h_{rd}|\sqrt{P_r^m}\right)^2\bigg), \end{aligned} \tag{8}$$

where $C_1$ represents the rate of the common part that can be decoded at $\mathcal{R}$, $C_2$ the private part that can be decoded at $\mathcal{D}$ provided the common part has been decoded correctly, and $C_3$ both the common and private parts that can be jointly decoded at $\mathcal{D}$. These rates are achievable provided the standard full channel knowledge at receivers and the source-relay coherent phase knowledge discussed earlier.

The PDF relaying scheme can be deployed using a cooperation policy $E$ that determines whether $\mathcal{S}$ should decide to exploit the help of $\mathcal{R}$ to communicate a message to $\mathcal{D}$, or directly convey its message to $\mathcal{D}$ depending on the relative quality of the $\mathcal{S}$-to-$\mathcal{R}$ and $\mathcal{S}$-to-$\mathcal{D}$ links. Subsequently, the average rate in this deployment can be obtained as follows

$$R = \mathbb{E}\{\min(C_1 + C_2, C_3)|E\}\mathbb{P}\{E\} + \mathbb{E}\{C|\bar{E}\}\mathbb{P}\{\bar{E}\} \tag{9}$$

where $C$ is the direct transmission capacity and $\bar{E}$ is the complementary event of $E$ with $\mathbb{P}\{\bar{E}\} = 1 - \mathbb{P}\{E\}$. In Section IV, we will discuss cooperation policies in detail.

## III. Cellular Network Geometry and User-Assisted Relaying Deployment

In this section, we provide an overview of the cellular system under consideration and describe the stochastic geometry network model. We then describe the channel model and the received signals when deploying PDF user-assisted relaying in the whole network. Finally, we develop the out-of-cell interference model and map the channel model with interference back to the standard form for subsequent analysis.

### A. Network Geometry Model

We consider a cellular system which consists of multiple cells, each cell has a single base station and each base station serves multiple users. Each of the users uses a distinct resource block, subsequently, no intra-cell interference is present. The only interference that affects each user is the out-of-cell interference due to frequency reuse in all other cells. We assume that each user is served by the single base station that is closest to that user. Within this system, we study the impact on performance of the cooperation technique in which a user can relay its message to the base station through the closest user that is in an idle state in addition to the direct link using the PDF relaying scheme described in Section II.

For the geometric model, we employ stochastic geometry to describe the uplink cellular network as shown in Fig. 2. We assume that the active users in different cells that will contend for the same resource block and cause interference to each other are distributed on a two-dimensional plane according to a homogeneous and stationary Poisson point process (PPP) $\Phi_1$ with intensity $\lambda_1$. We also assume that $\Phi_1$ is independent of another PPP $\Phi_2$ with intensity $\lambda_2$ that represents the distribution of another set of UEs that are in an idle state and can participate in relaying the messages transmitted by UEs in $\Phi_1$. Furthermore, under the assumption that each BS serves a single mobile in a given resource block, we follow the same approach in describing BSs distribution as proposed in [23], where each BS is uniformly distributed in the Voronoi cell of its served UE. In Fig. 3, we present an example layout of the proposed model for the uplink user-assisted relaying cellular network.

### B. User-Assisted Relaying Deployment Channel Model

Given the PDF relaying scheme in Section II, we now describe its channel model when deployed in a network. The only difference between the stand-alone channel model in Section II and the network channel model introduced in this section is that here we consider out-of-cell interference, i.e., the interference from all other cells to the $i^{th}$ cell due to frequency reuse.

In the relaying case, we model received signals at the $i^{th}$ relay and the destination, respectively, during the first phase as

$$\begin{aligned} Y_{r,i}^b &= h_{sr}^{(i)} x_{s,i}^b + I_{r,i}^b + Z_{r,i}^b, \\ Y_{d,i}^b &= h_{sd}^{(i)} x_{s,i}^b + I_{d,i}^b + Z_{d,i}^b \end{aligned} \tag{10}$$

where $I_{r,i}^b$ and $I_{d,i}^b$ represents the interference received at the $i^{th}$ relay and destination.

Also, we model the received signal at the $i^{th}$ destination during the second phase as

$$Y_{d,i}^m = h_{sd}^{(i)} x_{s,i}^m + h_{rd}^{(i)} x_{r,i}^m + I_{d,i}^m + Z_{d,i}^m \tag{11}$$

where $I_{d,i}^m$ represents the interference received at the $i^{th}$ destination.

In the direct transmission case, we model the received signal at the $i^{th}$ destination as

$$Y_{d,i} = h_{sd}^{(i)} x_{s,i} + I_{d,i} + Z_{d,i} \tag{12}$$

where $I_{d,i}$ represents the average interference received at the $i^{th}$ destination during the whole two-phase transmission period.

Note that in our model, we do not use power control at each user but assume the worst-case scenario where each user is transmitting at the maximum allowable power. Thus our analysis results should represent a lower bound on the actual system performance, in the sense that the actual out-of-cell interference will likely be less than what we derive in our subsequent analysis.

### C. Out-of-Cell Interference

To develop the interference model, we assume perfect frame synchronization which can be justified by the fact that LTE-Advanced imposes very strict requirements on time synchronization. Failure to comply with the synchronization requirements impacts the performance of the various features developed in the standard such as LTE-A Coordinated multi-Point (CoMP) and enhanced Inter-Cell Interference Coordination (e-ICIC) [3], [24], [25]. Carrier networks achieve the necessary precision and accuracy in synchronization based on a very precise and accurate primary reference which is mainly obtained by signals transmitted by GNSS satellite systems.

Interference can therefore be expressed at the destination, during the first and second phase, and at the relay during the first phase, respectively, as follows

$$\begin{aligned} I_{d,i}^{b} &= \sum_{k\neq i} B_k h_{sd}^{(k,i)} x_{s,k}^{b} + (1-B_k) h_{sd}^{(k,i)} x_{s,k}, \\ I_{d,i}^{m} &= \sum_{k\neq i} B_k \left( h_{sd}^{(k,i)} x_{s,k}^{m} + h_{rd}^{(k,i)} x_{r,k}^{m} \right) \\ &\quad + (1-B_k) h_{sd}^{(k,i)} x_{s,k}, \\ I_{r,i}^{b} &= \sum_{k\neq i} B_k h_{sr}^{(k,i)} x_{s,k}^{b} + (1-B_k) h_{sr}^{(k,i)} x_{s,k}, \end{aligned} \quad (13)$$

where the summation is over all the active users. In the direct transmission case, the interference term $I_{d,i}$ in eq. (12) is equivalent to $I_{d,i}^{b}$ during the first phase and $I_{d,i}^{m}$ during the second phase. Note that the interference at the destination BS and relay user during the first phase results only from the active users (sources) in either cooperation or direct transmission mode, and the interference at the destination during the second phase results from both the active and relaying users if in cooperation mode or the active users if in direct transmission mode. Here, $h_{sd}^{(k,i)}$ and $h_{rd}^{(k,i)}$, respectively, are the channel fading from the $k^{th}$ active UE in $\Phi_1$ and the associated relaying UE in $\Phi_2$ to the BS associated with the $i^{th}$ active UE in $\Phi_1$; and $h_{sr}^{(k,i)}$ is the channel fading from the $k^{th}$ active UE in $\Phi_1$ to the relaying UE associated with the $i^{th}$ active UE in $\Phi_1$.

The Bernoulli random variable $B_k \sim$ Bern $(\rho_1)$ captures the transmission strategy of the $k^{th}$ UE in $\Phi_1$ with success probability $\rho_1$, where $B_k = 1$ is used to indicate the $k^{th}$ active UE decision to exploit the help of another idle UE and apply the relaying transmission strategy, and $B_k = 0$ indicates direct transmission. A Bernoulli random variable can represent the transmission strategy with a certain probability, $\rho_1$, because, as we show in Section IV, the developed cooperation policies will be independent for each active user. We derive the cooperation probability $\rho_1$ for the different policies later in that section.

It is worth noting that the perfect frame synchronization assumption can be relaxed to assume only transmission phase synchronization by incorporating another Bernoulli random variable as in [26]. In such phase synchronization, a user can be in phase 1 while another user can be in phase 2, as long as these transmission phases are time-synchronized at the beginning.

For a given setting of nodes locations, based on the interference model in Eq. (13), we can use the fact that interference at either the relay or destination is the sum of an infinite number of signals undergoing independent fading from nodes distributed in the infinite 2-$D$ plane and invoke the law of large numbers to approximate the interference as a complex Gaussian distribution. Also, since the transmitted codewords are realization of a complex Gaussian with zero mean, it is justified to set the mean of interference to zero. To fully characterize interference as a complex Gaussian distribution, we define their distributions independently as $I_{d,i}^{b} \sim \mathcal{CN}(0, \mathcal{Q}_{d,i}^{b})$, $I_{d,i}^{m} \sim \mathcal{CN}(0, \mathcal{Q}_{d,i}^{m})$, and $I_{r,i}^{b} \sim \mathcal{CN}(0, \mathcal{Q}_{r,i})$, with the variances derived later in Section V. The power of these interference terms which correspond to the variance of the Gaussian random variables are function of node locations and hence vary with different network realizations.

Note that in our analysis, we implicitly assume no temporal or spatial correlation for the interferences at the relay UE and the BS, as noted in [27]. Temporal correlation does not arise since in our model, the channels are independent from frame to frame and all transmitted signals are independent from phase to phase. Spatial correlation can potentially arise when the BS and relay UE are closely or co-located, but in practice this does not occur since the BS is usually much higher than the UEs and we only need a separation of half a wavelength to achieve spatial independence. Further, as we will see in Section VI, user-assisted relaying deployment is most effective for active users towards the cell edge with a relay UE approximately midway between the active UE and the BS, making the issue of spatial correlation irrelevant.

### D. Equivalent Standard Channel Model

Given the channel model with interference in Eqs. (10)−(12) and the interference model discussed in Section III-C, we can convert the channel model in case of relaying into the standard form of Eqs. (1)−(2) to capture the effects of interference into the channel fading as

$$\begin{aligned} \tilde{Y}_{r,i}^{b} &= \tilde{h}_{sr}^{(i)} x_{s,i}^{b} + \tilde{Z}_{r,i}^{b}, \\ \tilde{Y}_{d,i}^{b} &= \tilde{h}_{sd}^{(b,i)} x_{s,i}^{b} + \tilde{Z}_{d,i}^{b}, \\ \tilde{Y}_{d,i}^{m} &= \tilde{h}_{sd}^{(m,i)} x_{s,i}^{m} + \tilde{h}_{rd}^{(i)} x_{r,i}^{m} + \tilde{Z}_{d,i}^{m}, \end{aligned} \quad (14)$$

where the new channel fading terms are defined as

$$\begin{aligned} \tilde{h}_{sr}^{(i)} &= \frac{h_{sr}^{(i)}}{\sqrt{\mathcal{Q}_{r,i} + \sigma^2}}, \quad \tilde{h}_{sd}^{(b,i)} = \frac{h_{sd}^{(i)}}{\sqrt{\mathcal{Q}_{d,i}^{b} + \sigma^2}}, \\ \tilde{h}_{sd}^{(m,i)} &= \frac{h_{sd}^{(i)}}{\sqrt{\mathcal{Q}_{d,i}^{m} + \sigma^2}}, \quad \tilde{h}_{rd}^{(i)} = \frac{h_{rd}^{(i)}}{\sqrt{\mathcal{Q}_{d,i}^{m} + \sigma^2}}, \end{aligned} \quad (15)$$

and the equivalent noise terms $\tilde{Z}_{r,i}^{b}, \tilde{Z}_{d,i}^{b}$, and $\tilde{Z}_{d,i}^{m}$ are now all i.i.d $\mathcal{CN}(0,1)$. Using these equivalent standard channels, we can compute the relaying transmission rates as in Eqs. (7)−(9). Note that the standard assumption of full channel knowledge at receivers still holds, since the $\mathcal{Q}$ terms represent the interference powers which can be measured on reception. The coherent source-relay phase knowledge assumption is unaffected.

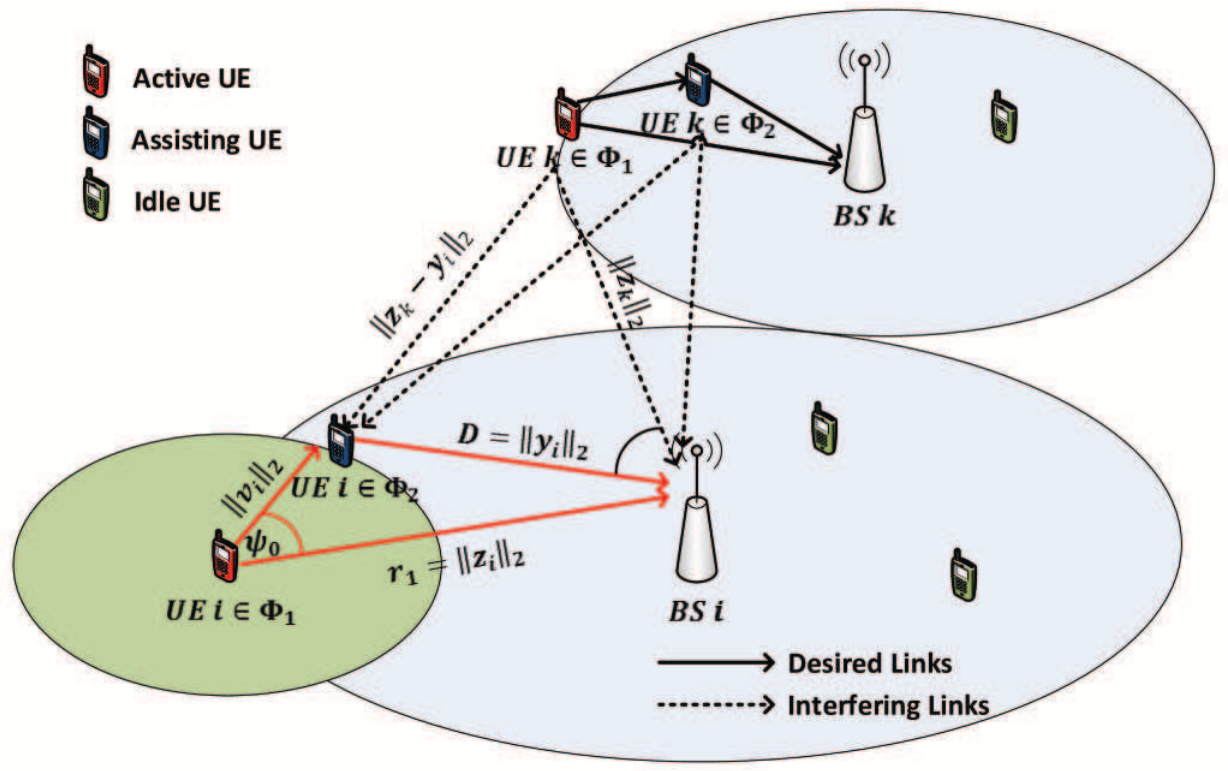

Fig. 2: System model of uplink user-assisted relaying.

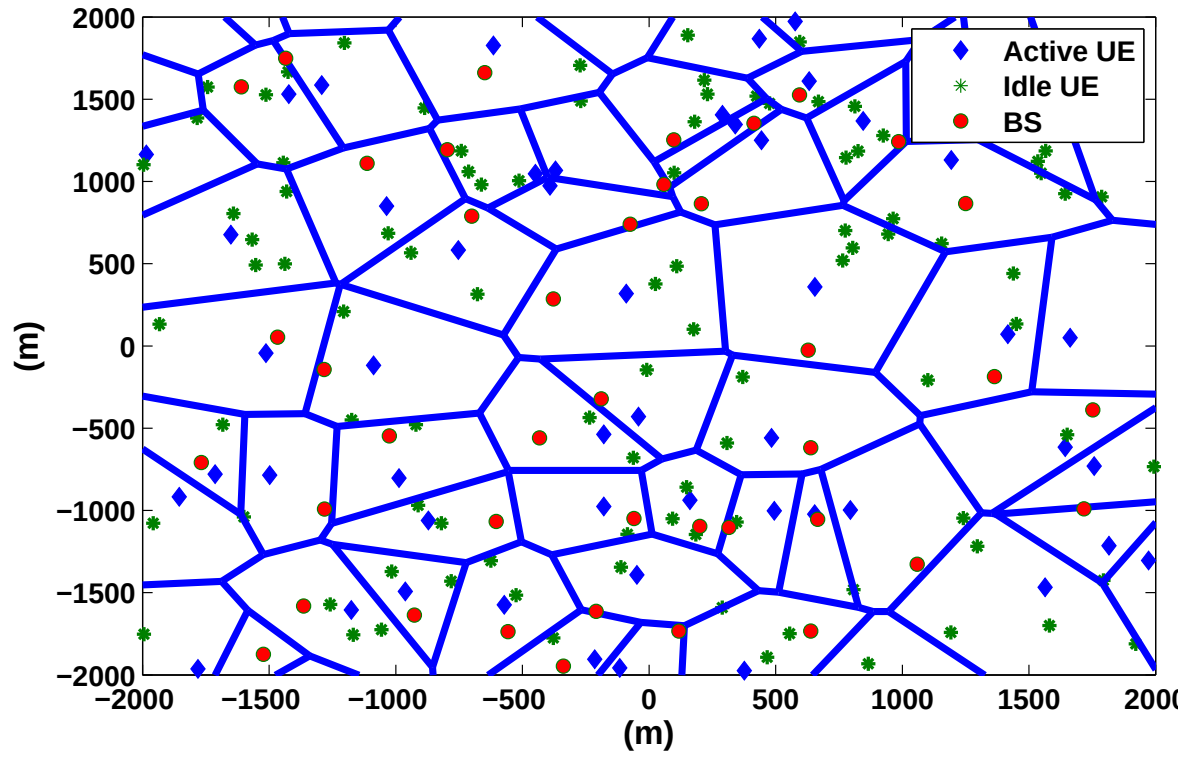

Fig. 3: Sample layout of a cellular network using Stochastic Geometry ($\lambda_2 = 2\lambda_1$).

Similarly, in the direct transmission case, the received signal at the $i^{th}$ destination can be remodeled into the form of Eq. (3) with the new channel fading assumed to be the same as $\tilde{h}_{sd}^{(b,i)}$ during the transmission time of the first phase and $\tilde{h}_{sd}^{(m,i)}$ during the second phase. Hence, the achievable rate of the direct transmission case can be obtained as

$$C = \alpha_1 \log\left(1 + \left|\tilde{h}_{sd}^{(b,i)}\right|^2 P_{s,i}\right) + \alpha_2 \log\left(1 + \left|\tilde{h}_{sd}^{(m,i)}\right|^2 P_{s,i}\right). \tag{16}$$

Channel gains can be further split into small scale and path loss fading components as

$$\left|h_{sd}^{(i)}\right|^2 = g_{sd}^{(i)} \|\mathbf{z}_i\|_2^{-\alpha}, \qquad \left|h_{sr}^{(i)}\right|^2 = g_{sr}^{(i)} \|\mathbf{v}_i\|_2^{-\alpha},$$

$$\left|h_{rd}^{(i)}\right|^2 = g_{rd}^{(i)} D^{-\alpha}, \qquad \left|h_{sd}^{(k,i)}\right|^2 = g_{sd}^{(k,i)} \|\mathbf{z}_k\|_2^{-\alpha}, \tag{17}$$

$$\left|h_{sr}^{(k,i)}\right|^2 = g_{sr}^{(k,i)} \|\mathbf{z}_k - \mathbf{y}_i\|_2^{-\alpha}, \qquad \left|h_{rd}^{(k,i)}\right|^2 = g_{rd}^{(k,i)} \|\mathbf{z}_k\|_2^{-\alpha}, \tag{18}$$

with
$$D^2 = \|\mathbf{z}_i\|^2 + \|\mathbf{v}_i\|^2 - 2\|\mathbf{z}_i\|\|\mathbf{v}_i\| \cos\psi_0, \tag{19}$$

where we use the law of cosines to obtain eq. (19); $\mathbf{z}_k$ and $\mathbf{v}_i$ are vectors representing the 2-$D$ locations of the UEs in $\Phi_1$ and $\Phi_2$, respectively; and $\psi_0 \sim \mathcal{U}[0:2\pi]$ is a uniform random variable that represents the angle between the two vectors $\mathbf{z}_i$ and $\mathbf{y}_i$ connecting the $i^{th}$ UE to its base station and relaying node. $g_{sd}^{(k,i)}$, $g_{rd}^{(k,i)}$, and $g_{sr}^{(k,i)}$ are all i.i.d. $\sim \exp(1)$ and represent the power gain of the small scale Rayleigh fading channels.

Note that in Eq. (18), we use the out-of-cell interference far field approximation and set the location of out-of-cell source interferer and its associated relay to be the same as also done for CoMP in [13]. The results in [13] were shown to have an analytic performance similar to that of simulation without the far field approximation in the case of non-cooperating base station, but this similarity does not completely hold in case of full cooperation between neighboring base stations. The case is different in our model since, as discussed in Section IV next, we restrict the selection of each active UE for its relay to be the closest idle UE that is located within the coverage of their base station. Hence, the expected distance between the cooperating nodes is much less than that between neighboring base stations, justifying the far field approximation.

## IV. Cooperation Policies and Probability

In this section, we discuss three cooperation policies: an ideal policy $E_1$ based on the stand-alone policy in [18], a new pure geometric policy $E_2$ and a new hybrid policy $E_3$ that defines whether a UE should select the relaying strategy or the direct transmission strategy. Then, we develop analytic cooperation probability expressions for the two new and more practical policies. The decision making nodes according to these policies can either be the base station only or both the active user and the closest idle user associated with it as a relay, but we leave this choice up to the specific implementation.

### A. Cooperation Policies Definition

We first discuss the ideal cooperation policy, $E_1$, which requires the decision making nodes to know the instantaneous channels between the active user and the associated relaying user as well as its serving base station. It also requires full knowledge of the interference at the relay user and base station on decision making. Effectively, the ideal policy relies on knowledge of the instantaneous SINRs of the relay link and the direct link at the decision making node. This policy is defined using the equivalent standard channel model in Section III-D as [18]

$$E_1 = \left\{\left|\tilde{h}_{sr}^{(k)}\right|^2 \geq \left|\tilde{h}_{sd}^{(k)}\right|^2\right\} \tag{20}$$

$$\simeq \left\{\frac{g_{sr} r_2^{-\alpha}}{\mathcal{Q}_{r,k}} \geq \frac{g_{sd} r_1^{-\alpha}}{\mathcal{Q}_{d,k}^b}\right\} \tag{21}$$

where $r_1 = \|\mathbf{z}_i\|_2$ and $r_2 = \|\mathbf{v}_i\|_2$ denote the direct distance between $\mathcal{S}$ and $\mathcal{D}$ and cooperation distance between $\mathcal{S}$ and $\mathcal{R}$, respectively, as shown in Fig. 2. This event $E_1$ identifies whether an idle UE will be associated as a relay for the $k^{th}$ UE and participate in transmission.

Next, we propose two more practical policies that require less information in decision making. In an interference limited scenario, we can ignore the effect of the noise variance $\sigma^2$ and hence follows Eq. (21) from which we can conclude that when the interference powers at the relay and at the destination are approximately the same, the cooperation event depends mainly on the distances from the source to relay and to the destination. Taking this fact into account and averaging out the effects of small scale fading, we propose a more practical, pure geometric cooperation policy, $E_2$, as an approximate to the cooperation event in Eq. (21) by assuming independence between the cooperation policy and interference, i.e., the interference terms $I^b_{d,k}$, $I^m_{d,k}$ and $I^b_{r,k}$ will not affect the local cooperation decision of the $k^{th}$ UE in $\Phi_1$.

The pure geometric policy $E_2$ is defined as

$$E_2 = \{r_2 \leq r_1, D \leq r_1\}. \tag{22}$$

Policy $E_2$ is more practical than policy $E_1$ in the sense that it does not require full knowledge of both the channel fading and the interference at the decision making node. Instead, it only requires the decision making nodes to know the distances from the active user to the nearest idle user and to the base station. It represents a practical decision making strategy for fast fading channels, requiring no knowledge of the channel fading. The extra condition $D \leq r_1$ ensures that the relaying user is closer to the base station than the active user, effectively eliminating all cases that can result in an infinite interference at the relay. Thus policy $E_2$ only chooses the relaying users approximately within half a circle of radius $r_1$ centering at the base station.

To realize policy $E_2$, different network aided positioning techniques can be used by the base station to obtain information about UEs locations as surveyed in [28]. For mobile users, a location tracking algorithm based on the Kalman filter with velocity estimation and direction finder is proposed in [29]. This algorithm is used in [30] to develop an interference coordination algorithm in a D2D communications underlaying cellular networks.

The last policy, $E_3$, is proposed for slow fading channels where small scale fading parameters estimation and their feedback to the decision making node is feasible. We denote this policy as the hybrid fading and geometric policy and define it as

$$E_3 = \{g_{sd} r_1^{-\alpha} \leq g_{sr} r_2^{-\alpha}, D \leq r_1\}. \tag{23}$$

Note that this cooperation policy is still independent from the interference as in the pure geometric cooperation policy $E_2$.

Note that the ideal policy $E_1$ may result in an event where the relaying user belongs to a different cell than that of the active user, where the connection between the two nodes is still supported by D2D communication. Our proposed policies $E_2$ and $E_3$, however, have conditions to circumvent such an event. Specifically, this event is unlikely in $E_2$ by ensuring that $D \leq r_1$ and $r_2 \leq r_1$, diminishing the chance of selecting an out-of-cell relaying user. In fact, this chance is zero for circular cells as considered for the cell under study in Section V. In $E_3$, there is a small probability that this event occurs, but this probability is also limited by $D \leq r_1$. In both policies, choosing an out-of-cell relay has a minimal impact on performance, as confirmed later by our analysis and simulation results. Moreover, choosing an out-of-cell relaying user can be an issue only in theoretical analysis because of the use of stochastic geometry in which interference at the relaying user could be infinite. In practice, however, out-of-cell relaying user is feasible over D2D links and suffers only finite interference from the active user in its own cell.

### B. Cooperation Probabilities

Here we analytically derive the cooperation probabilities for the proposed geometric policy $E_2$ and the hybrid policy $E_3$. For the ideal policy $E_1$, analytic evaluation of the cooperation probability is rather complicated because of the inter-dependency between the cooperation decision and consequential interference among different cells, hence we use numerical simulations instead. We perform the analysis at a random BS assuming that it is associated with the $i^{th}$ UE. We assume that randomly picking this BS is equivalent to selecting a point uniformly distributed in the $\mathbb{R}^2$ plane as done in [23]. Under this assumption, the distribution of the distance $r_1$ between the $i^{th}$ UE and its associated BS can be shown to be Rayleigh distributed directly from the null probability of a two dimensional PPP distribution.

Moreover, we can assume due to the stationarity of the PPP and the independence of $\Phi_2$ from BSs distribution that the location of the UE associated with the BS under study represents the origin (typical) point of $\Phi_2$. Then, each UE in $\Phi_1$ chooses the closest UE in $\Phi_2$ to assist it in relaying its message to the serving BS. Hence, similar to source-to-destination distance, the distribution of the source-to-relay distance $r_2$ between the $i^{th}$ UE and its associated relaying UE can be also shown to be Rayleigh distributed directly from the null probability of a two dimensional PPP. Thus, we have

$$\begin{aligned} f_{r_1}(r_1) &= 2\pi\lambda_1 r_1 e^{-\lambda_1 \pi r_1^2}, \\ f_{r_2}(r_2) &= 2\pi\lambda_2 r_2 e^{-\lambda_2 \pi r_2^2}. \end{aligned} \tag{24}$$

The probabilities of cooperation policies $E_2$ and $E_3$ in (22) and (23) denoted as $\rho_2$ and $\rho_3$, respectively, which can be approximations of the cooperation probability $\rho_1$ of $E_1$ in Eq. (21) are derived in closed form as in Theorem 1.

**Theorem 1** (Cooperation probabilities)**.** *The probability of deploying user-assisted relaying for a randomly located active user within a cell can be evaluated as follows:*

(i) *For policy $E_2$*

$$\begin{aligned} \rho_2 = &\int_{-\pi/2}^{-\pi/3} \frac{2\lambda_2 \cos^2\psi_0}{\pi(\lambda_1 + 4\lambda_2 \cos^2\psi_0)} d\psi_0 \\ &+ \int_{\pi/3}^{\pi/2} \frac{2\lambda_2 \cos^2\psi_0}{\pi(\lambda_1 + 4\lambda_2 \cos^2\psi_0)} d\psi_0 + \frac{\lambda_2}{3(\lambda_1 + \lambda_2)}. \end{aligned} \tag{25}$$

$$\mathcal{Q}^b_{d,i} = \sum_{k\neq i} B_k \left|h^{(k,i)}_{sd}\right|^2 P^b_{s,k} + (1-B_k)\left|h^{(k,i)}_{sd}\right|^2 P_{s,k}. \tag{29}$$

$$\mathcal{Q}^m_{d,i} = \sum_{k\neq i} B_k \left(\left|h^{(k,i)}_{sd}\right|^2 P^m_{s,k} + \left|h^{(k,i)}_{rd}\right|^2 P^m_{r,k} + 2\left|h^{(k,i)}_{sd}\right|\left|h^{(k,i)}_{rd}\right|\sqrt{P^{m1}_{s,k}P^m_{r,k}}\cos\left(\theta_{k2,i}-\theta_{kr,i}\right)\right) + (1-B_k)\left|h^{(k,i)}_{sd}\right|^2 P_{s,k}. \tag{30}$$

$$\mathcal{Q}^m_{d,i} = \sum_{k\neq i} \left[B_k \left(\left|h^{(k,i)}_{sd}\right|^2 P^m_{s,k} + \left|h^{(k,i)}_{rd}\right|^2 P^m_{r,k}\right)\right] + (1-B_k)\left|h^{(k,i)}_{sd}\right|^2 P_{s,k}. \tag{31}$$

$$\mathcal{Q}_{r,i} = \sum_{k\neq i} B_k \left|h^{(k,i)}_{sr}\right|^2 P^b_{s,k} + (1-B_k)\left|h^{(k,i)}_{sr}\right|^2 P_{s,k}. \tag{32}$$

(ii) *For policy $E_3$*

$$\rho_3 = \int_0^2 f_\beta(z)\int_{-\pi/2}^{-\cos^{-1}(z/2)} \frac{2\lambda_2\cos^2\psi_0}{\pi(\lambda_1+4\lambda_2\cos^2\psi_0)}d\psi_0 dz + \int_0^2 f_\beta(z)\int_{\cos^{-1}(z/2)}^{\pi/2} \frac{2\lambda_2\cos^2\psi_0}{\pi(\lambda_1+4\lambda_2\cos^2\psi_0)}d\psi_0 dz + \int_0^2 f_\beta(z)\frac{\lambda_2\cos^{-1}(z/2)}{\pi\left(\lambda_1+\lambda_2\right)}dz + \int_2^\infty\int_{-\pi/2}^{\pi/2} f_\beta(z)\frac{2\lambda_2\cos^2\psi_0}{\pi(\lambda_1+4\lambda_2\cos^2\psi_0)}d\psi_0 dz, \tag{26}$$

*where $\beta = \left(\frac{g_{sr}}{g_{sd}}\right)^{1/\alpha}$ and $f_\beta(z)$ is the probability density function (PDF) of $\beta$ defined as*

$$f_\beta(z) = \frac{\alpha z^{\alpha-1}}{\left(1+z^\alpha\right)^2}. \tag{27}$$

*Proof:* See Appendix A for details. ■

We use numerical integration to compare between these probabilities in Section VI-A. We can note by investigating equations (25) and (26) that they are both proportional to the users density ratio $\lambda_2/\lambda_1$ and that both probabilities $\rho_2$ and $\rho_3$ achieve their maximum when this ratio approaches infinity, i.e. as the density of the idle users increases. We can evaluate the maximum probability achieved by both cooperation policies $E_2$ and $E_3$ as

$$\rho_2^{max} = \lim_{\lambda_2\to\infty}\rho_2(\lambda_1,\lambda_2) = 0.5,$$
$$\rho_3^{max} = \lim_{\lambda_2\to\infty}\rho_3(\lambda_1,\lambda_2) = 0.5. \tag{28}$$

The maximum probabilities achieved of 0.5 is because of our restriction of the spatial cooperation domain to the idle UEs that are closer to the BS than the active user seeking cooperation. Effectively, we only consider potential relays approximately in a half circle centered at the active user and inside the cell under consideration.

## V. Out-of-Cell Interference Analysis

User-assisted relaying actually increases the amount of out-of-cell interference in the network as some idle users are now transmitting when relaying information of active users. It is therefore necessary to understand this out-of-cell interference power, particularly its distribution, in order to assess the overall impact of user-assisted relaying on system performance. Given the stochastic geometry system model described in Section III-A with PPP distributions, tools from stochastic geometry can be used to analytically derive the moments of the interference power in the network.

To study the performance of user assisted relaying, we consider a cell with fixed radius, $R_c$, and analyze the out-of-cell interference to this cell as a typical case for the network. The radius $R_c$ is typically proportional to the active users density as $R_c = 1/(2\sqrt{\lambda_1})$, but we keep $R_c$ as a parameter in our analysis and later investigate different values in numerical results and simulations. Since it is difficult to describe the exact distribution of out-of-cell interference power, here we choose to model the interference power to the cell under study as a Gamma distribution by fitting the first two moments of the interference power analytically developed using stochastic geometry of the field of interferers outside that cell. The fit of this Gamma distribution model will be evaluated in Section VI to assess the impact of the model on the accuracy of system performance.

Having an analytical interference model can significantly simplify system performance analysis by removing the need for time and labor intensive simulation. Further, such an analytical interference model also allows tractable performance analysis with detailed understanding of the impact of each parameter which may not be feasibly obtained by simulation.

### A. Analytic Development of Moments of the Interference Power

We start by deriving interference power moments, for which, we first build up from the interference expressions in Eq. (13) to develop the interference power at the $i^{th}$ destination BS during the first and second phase, respectively, as in Eq. (29) and Eq. (30). Similarly, interference power at the idle UE associated as a relay with the $i^{th}$ active UE can be written as in Eq. (32). Here, $\theta_{k2,i}$ and $\theta_{kr,i}$ are realizations of independent and uniformly distributed random angle variables. We use the approximation described in [13] where interference is averaged over the reception angles and since $\mathbb{E}_{\theta_{k2,i},\theta_{kr,i}}\left[\cos\left(\theta_{k2,i}-\theta_{kr,i}\right)\right]=0$, we can rewrite the interference power at the $i^{th}$ destination BS during the second phase as in Eq. (31). This approximation implies that even though the out-of-cell source-relay pair transmits coherently (i.e. beamform) to its own destination, the two signals go through different channels to the cell under consideration and appear independent of each other.

Next, we use the *Laplace* transform of interference power at the $i^{th}$ destination BS during the first and second phase as derived in Appendix B to characterize the moments of interference.

**Theorem 2** (Interference Power *Laplace* Transform)**.** *To characterize the moments of interference power, its Laplace transform is derived as follows:*

*(i) At the $i^{th}$ destination BS during the first and second phase, respectively*

$$\mathcal{L}_{\mathcal{Q}^b_{d,i}}(s)=\exp\left(-2\pi\lambda_1\int_{R_c}^{\infty}\left(1-\mathcal{L}_{\mathcal{J}^b_{d,i}}(s,r)\right)rdr\right), \quad (33)$$

$$\mathcal{L}_{\mathcal{Q}^m_{d,i}}(s)=\exp\left(-2\pi\lambda_1\int_{R_c}^{\infty}\left(1-\mathcal{L}_{\mathcal{J}^m_{d,i}}(s,r)\right)rdr\right), \quad (34)$$

*where $\mathcal{L}_{\mathcal{J}^b_{d,i}}(s,\|\mathbf{z}_k\|_2)$ and $\mathcal{L}_{\mathcal{J}^m_{d,i}}(s,\|\mathbf{z}_k\|_2)$ are expressed as*

$$\begin{aligned}\mathcal{L}_{\mathcal{J}^b_{d,i}}(s,\|\mathbf{z}_k\|_2) &= \rho_1\mathcal{L}_G\left(s\|\mathbf{z}_k\|_2^{-\alpha}P^b_{s,k}\right)\\ &\quad+(1-\rho_1)\mathcal{L}_G\left(s\|\mathbf{z}_k\|_2^{-\alpha}P_{s,k}\right), \quad (35)\\ \mathcal{L}_{\mathcal{J}^m_{d,i}}(s,\|\mathbf{z}_k\|_2) &= \rho_1\mathcal{L}_G\left(s\|\mathbf{z}_k\|_2^{-\alpha}P^m_{s,k}\right)\mathcal{L}_G\left(s\|\mathbf{z}_k\|_2^{-\alpha}P^m_{r,k}\right)\\ &\quad+(1-\rho_1)\mathcal{L}_G\left(s\|\mathbf{z}_k\|_2^{-\alpha}P_{s,k}\right). \quad (36)\end{aligned}$$

*where $\mathcal{L}_G(s)=\frac{1}{1+s}$ is the Laplace transform of an exponential random variable $G\sim\exp(1)$.*

*(ii) At the idle UE associated as a relay for the $i^{th}$ active UE*

$$\mathcal{L}_{\mathcal{Q}_{r,i}}(s)=\exp\left(-\lambda_1\int_0^{2\pi}\int_{R_c}^{\infty}\left(1-\mathcal{L}_{\mathcal{J}_{r,i}}(s,r,\theta)\right)rdrd\theta\right), \quad (37)$$

*where $\mathcal{L}_{\mathcal{J}_{r,i}}(s,\|\mathbf{z}_k\|_2,\theta_k)$ is defined as*

$$\begin{aligned}\mathcal{L}_{\mathcal{J}_{r,i}}(s,\|\mathbf{z}_k\|_2,\theta_k) &=\rho_1\mathcal{L}_G\left(s\|\mathbf{z}_k-\mathbf{y}_i\|_2^{-\alpha}P^b_{s,k}\right)\\ &\quad+(1-\rho_1)\mathcal{L}_G\left(s\|\mathbf{z}_k-\mathbf{y}_i\|_2^{-\alpha}P_{s,k}\right), \quad (38)\end{aligned}$$

*and the term $\|\mathbf{z}_k-\mathbf{y}_i\|_2$ can be written in terms of $\|\mathbf{z}_k\|_2$, distance $D$ defined in* (19) *and $\theta_k$ using the law of cosines as*

$$\|\mathbf{z}_k-\mathbf{y}_i\|_2^2=\|\mathbf{z}_k\|_2^2+D^2-2\|\mathbf{z}_k\|_2D\cos\theta_k. \quad (39)$$

*Proof:* See Appendix B for details. ■

**Lemma 1** (Interference Power Statistics)**.** *For network-wide deployment of user-assisted relaying, the out-of-cell interference generated at the destination BS and the relaying UE have the following statistics:*

*(i) The first two moments, mean and variance, of interference power at the destination BS during the $1^{st}$ and $2^{nd}$ phase, respectively, are*

$$\mathbb{E}\left[\mathcal{Q}^b_{d,i}\right]=\frac{2\pi\lambda_1\zeta_1}{\alpha-2}R_c^{2-\alpha},\qquad \mathbb{E}\left[\mathcal{Q}^m_{d,i}\right]=\frac{2\pi\lambda_1\zeta_3}{\alpha-2}R_c^{2-\alpha}, \quad (40)$$

$$var\left[\mathcal{Q}^b_{d,i}\right]=\frac{\pi\lambda_1\zeta_2}{\alpha-1}R_c^{2(1-\alpha)},\ var\left[\mathcal{Q}^m_{d,i}\right]=\frac{\pi\lambda_1\zeta_4}{\alpha-1}R_c^{2(1-\alpha)}. \quad (41)$$

*(ii) The first two moments, mean and variance, of interference power at the idle UE associated as a relay with the $i^{th}$ active UE are*

$$\mathbb{E}[\mathcal{Q}_{r,i}]=\lambda_1\zeta_1\int_0^{2\pi}\int_{R_c}^{\infty}\left(r^2+D^2-2rD\cos(\theta)\right)^{-\frac{\alpha}{2}}rdrd\theta, \quad (42)$$

$$var[\mathcal{Q}_{r,i}]=\lambda_1\zeta_2\int_0^{2\pi}\int_{R_c}^{\infty}\left(r^2+D^2-2rD\cos(\theta)\right)^{-\alpha}rdrd\theta, \quad (43)$$

*where*

$$\begin{aligned}\zeta_1 &= \rho_1P^b_{s,k}+(1-\rho_1)P_{s,k},\\ \zeta_2 &= 2\left[\rho_1(P^b_{s,k})^2+(1-\rho_1)P^2_{s,k}\right], \quad (44)\\ \zeta_3 &= \rho_1(P^m_{s,k}+P^m_{r,k})+(1-\rho_1)P_{s,k},\\ \zeta_4 &= 2\left[\rho_1(P^m_{s,k}+P^m_{r,k})^2+(1-\rho_1)P^2_{s,k}-\rho_1P^m_{s,k}P^m_{r,k}\right].\end{aligned}$$

*Proof:* The proof is straightforward by using the results of the interference power *Laplace* transform in Theorem 2 and evaluating the following formulas:

(i) At the destination

$$\begin{aligned}\mathbb{E}\left[\mathcal{Q}^b_{d,i}\right] &= -\frac{\partial\mathcal{L}_{\mathcal{Q}^b_{d,i}}(s)}{\partial s}\bigg|_{s=0},\\ \mathbb{E}\left[\mathcal{Q}^m_{d,i}\right] &= -\frac{\partial\mathcal{L}_{\mathcal{Q}^m_{d,i}}(s)}{\partial s}\bigg|_{s=0},\\ \text{var}\left[\mathcal{Q}^b_{d,i}\right] &= \frac{\partial^2\mathcal{L}_{\mathcal{Q}^b_{d,i}}(s)}{\partial s^2}\bigg|_{s=0}-\left(\mathbb{E}\left[\mathcal{Q}^b_{d,i}\right]\right)^2,\\ \text{var}\left[\mathcal{Q}^m_{d,i}\right] &= \frac{\partial^2\mathcal{L}_{\mathcal{Q}^m_{d,i}}(s)}{\partial s^2}\bigg|_{s=0}-\left(\mathbb{E}\left[\mathcal{Q}^m_{d,i}\right]\right)^2. \quad (45)\end{aligned}$$

(ii) At the relay

$$\begin{aligned}\mathbb{E}[\mathcal{Q}_{r,i}] &= -\frac{\partial\mathcal{L}_{\mathcal{Q}_{r,i}}(s)}{\partial s}\bigg|_{s=0}, \quad (46)\\ \text{var}[\mathcal{Q}_{r,i}] &= \frac{\partial^2\mathcal{L}_{\mathcal{Q}_{r,i}}(s)}{\partial s^2}\bigg|_{s=0}-\left(\mathbb{E}[\mathcal{Q}_{r,i}]\right)^2.\end{aligned}$$

■

Clearly from the above results for interference power statistics, the interference power is directly proportional to both the active users density, $\lambda_1$, and the transmission power levels represented by $\zeta_i, i\in[1:4]$ in Eq. (44), which agrees with intuition. Note also that the variance at the destination during the second phase is directly proportional to $\zeta_4$ which includes the full correlation term $P^m_{s,k}P^m_{r,k}$ between the $k^{th}$ active user and its associated relay transmission, even though the active user uses only part of its power during the second phase to coherently transmit with its associated relay. The above results further show the effect of the path loss exponent, $\alpha$, and cell radius, $R_c$, on the interference power statistics. For the practical cases when $\alpha\geq 2$, the interference power statistics are inversely proportional to the cell radius and approaches zero at the destination BS as the cell radius increases.

### B. Modeling Interference Power Distribution

A parameterized probability distribution, which includes a wide variety of curve shapes, is useful in the representation of data when the underlying model is unknown or difficult to obtain in closed form. A parameterized probability distribution is usually characterized by its flexibility, generality, and simplicity. Although distributions are not necessarily determined by their moments, the moments often provide useful information and are widely used in practice. For example, a four-parameter probability distribution is introduced in [31], [32] which is used to fit a set of data and match up to its fourth order moment.

In [33], the two-parameter Gamma distribution is used in a study of the downlink performance in a fixed-size cell within a cellular network to match the first two moments of a given distribution representing either the product of the small-scale and lognormal fading or the out-of-cell co-tier and cross-tier interference power distributions. It is shown that the Gamma distribution is a good approximation for the interference when the point under study is closer to the cell center, but fails to represent the actual interference distribution whenever the point under study is exactly at the cell edge. We use the same approach here and match a Gamma distribution to the first two moments of the interference power terms derived earlier in Lemma 1. In Section VI-B, we study the validity of this interference model in our network while varying parameters such as the locations of the active user and the associated relaying user, and the maximum transmit power allowed in the network.

The Gamma distribution is specified by two parameters, a shape parameter $k$ and a scale parameter $\theta$. Given a Gamma random variable $\gamma[k,\theta]$, its probability density function is defined as $F_{\gamma}(q|k,\theta) = \frac{q^{k-1}e^{-q/\theta}}{\theta^k\Gamma(k)}$, where the Gamma function $\Gamma(t)$ is defined as $\Gamma(t) = \int_0^\infty x^{t-1}e^{-x}dx$. The mean and variance of $\gamma[k,\theta]$ can be written in the following form [34]

$$\mathbb{E}[\gamma] = k\theta, \tag{47}$$

$$\text{var}[\gamma] = k\theta^2. \tag{48}$$

Given that we have analytically obtained the first two moments of the interference power terms of interest, we can estimate the shape and scale parameters of the Gamma distributed random variables $\gamma^b_{d,i}$, $\gamma^m_{d,i}$, and $\gamma^b_{r,i}$ that fit interference power terms $\mathcal{Q}^b_{d,i}$, $\mathcal{Q}^m_{d,i}$, and $\mathcal{Q}^b_{r,i}$, respectively, using the moments estimation described below in Lemma 2 as introduced in [33], [35].

**Lemma 2** (Estimation of Gamma Distribution Parameters using Moments Matching)**.** *Given a distribution $\mathcal{Q}_i$ with mean $\mathbb{E}[\mathcal{Q}_i]$ and variance var$[\mathcal{Q}_i]$, the shape and scale parameters $k_i$ and $\theta_i$ of the Gamma distributed random variable $\gamma_i[k_i,\theta_i]$ can be estimated as*

$$k_i = \frac{(\mathbb{E}[\mathcal{Q}_i])^2}{var[\mathcal{Q}_i]}, \qquad \theta_i = \frac{var[\mathcal{Q}_i]}{\mathbb{E}[\mathcal{Q}_i]}. \tag{49}$$

*Proof:* The proof follows easily, given that we know $\mathbb{E}[\mathcal{Q}_i]$ and var$[\mathcal{Q}_i]$, by equating the first two moments of the actual distribution random variable, $\mathcal{Q}_i$, to the moments of a Gamma distributed random variable, $\gamma_i[k_i,\theta_i]$, defined in (47) and (48). Then, using (47), we have $k_i = \mathbb{E}[\mathcal{Q}_i]/\theta_i$ and substituting into (48), we obtain the equations in (49). ■

## VI. Model Evaluation And Discussion

In this section, we use our geometric network model to numerically verify the validity of the analytical results and models established in the previous two sections. We first verify the validity of using the Rayleigh distribution to model both the cooperation (soure-to-relay) and direct (source-to-destination) distances. We then discuss and compare the different cooperation policies probabilities. Next, we evaluate and discuss the validity of using the Gamma distribution to model the out-of-cell interference power in a user-assisted relaying network. These validations confirm our analytical results and the fit of the Gamma distribution for out-of-cell interference power for the range of practical system parameters.

In both this section and Section VII next, to make a fair comparison between our analytical models and simulation, we assume in case of simulation that all cells other than the cell under study independently use the hybrid cooperation policy, $E_3$, whereas the cell under study uses the ideal cooperation policy, $E_1$. We use this simulation setting instead of the case of all cells using $E_1$ policy since the simulated setting gives more realistic performance and presents no inter-cell dependency between cooperation decision and interference, making the simulation computationally feasible.

### A. Geometric Model and Cooperation Probability

We present a sample layout of the network model for uplink user-assisted relaying in cellular system in Fig. 3 where $\lambda_2 = 2\lambda_1$. Both simulation and analytical results show that even for such a low density of idle users, there is a good probability to exploit cooperation in the network to help cell edge users. In Fig. 4, we first validate our choice of the Rayleigh distribution to model the distance of the source-to-relay link by generating around $5 \times 10^6$ sample network layouts using the stochastic geometry model where $\lambda_2 = 2$ is chosen for the results shown. Similarly, the validation of the source-to-destination distance distribution is shown in Fig. 5 for $\lambda_1 = 1/(16 \times 150^2)$ user per m$^2$. As Figs. 4 and 5 show, the Rayleigh distribution matches perfectly the simulated data.

In Fig. 6, we use numerical simulation to evaluate the probability of cooperation policy $E_3$ and compare it to the analytical closed-form probabilities obtained for both policies $E_2$ and $E_3$ in Eqs. (25)−(26). In simulation, we used policy $E_3$ where perfect knowledge of the small scale fading channel is assumed but not the out-of-cell interference. Based on Fig. 6, we can see that both cooperation policies have almost the same probability and they match closely the simulation results. These results show that even in the case of the least knowledge about the channel fading, policy $E_2$ can provide a similar cooperation result as other more complex policies requiring more information. The results suggest $E_2$ is suitable for practical deployment.

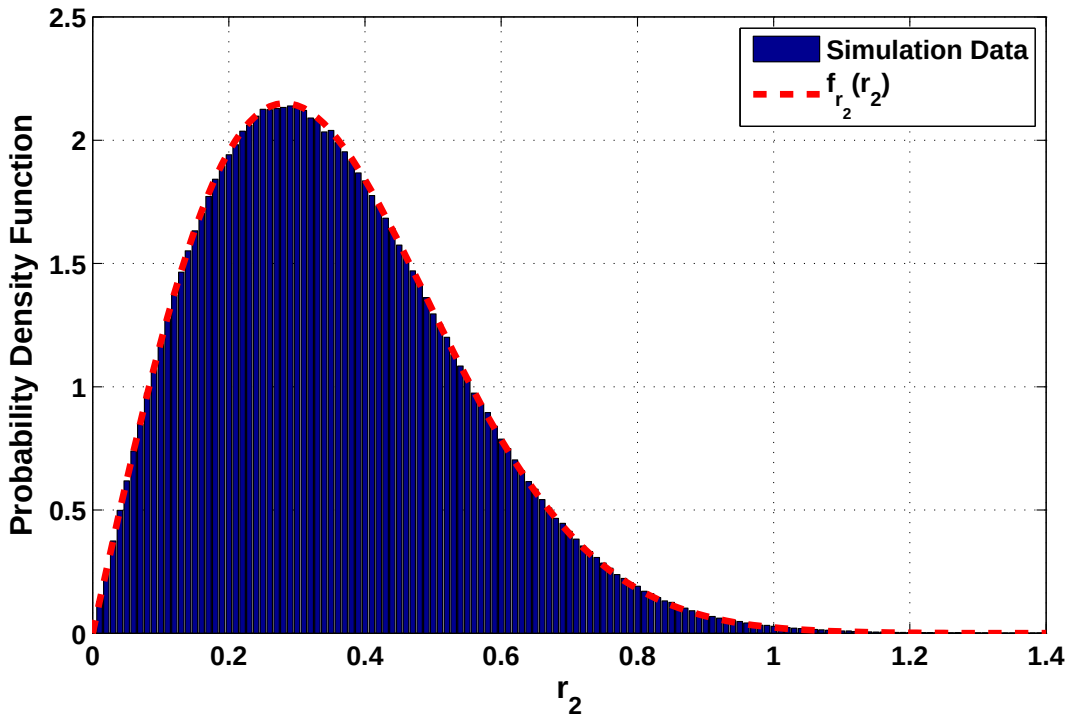

Fig. 4: Validation of the distribution of the cooperation link distance $r_2$ ($\lambda_2 = 2$).

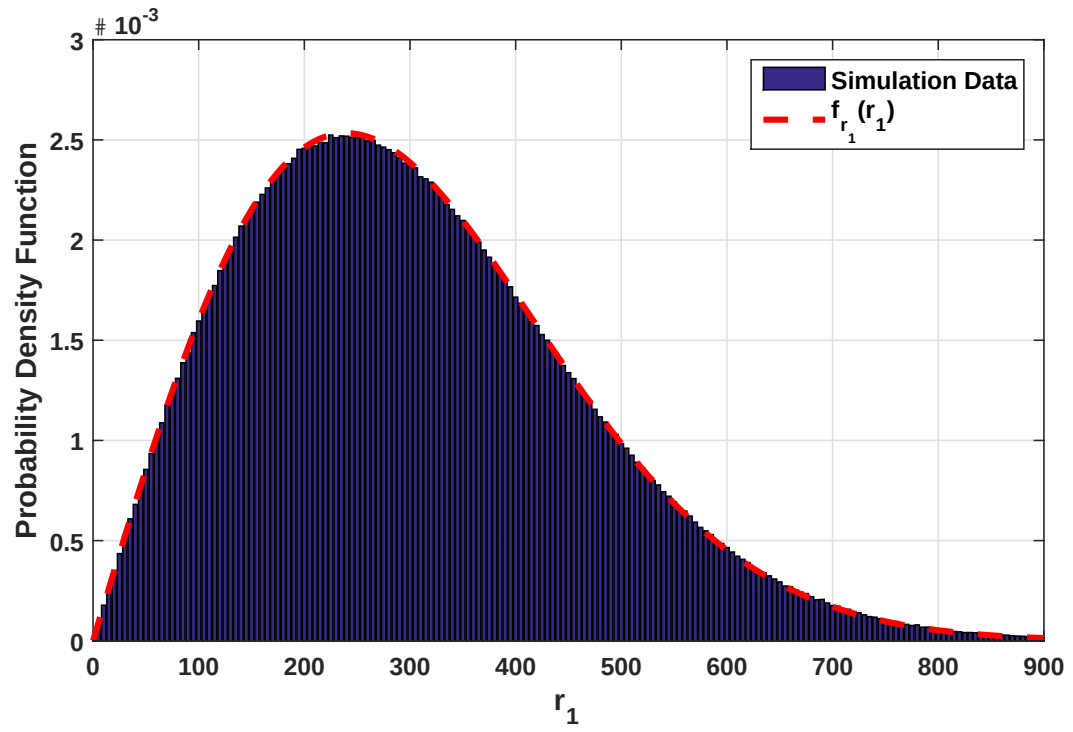

Fig. 5: Validation of the distribution of the direct link distance $r_1$ ($\lambda_1 = \frac{1}{16\times150^2}$).

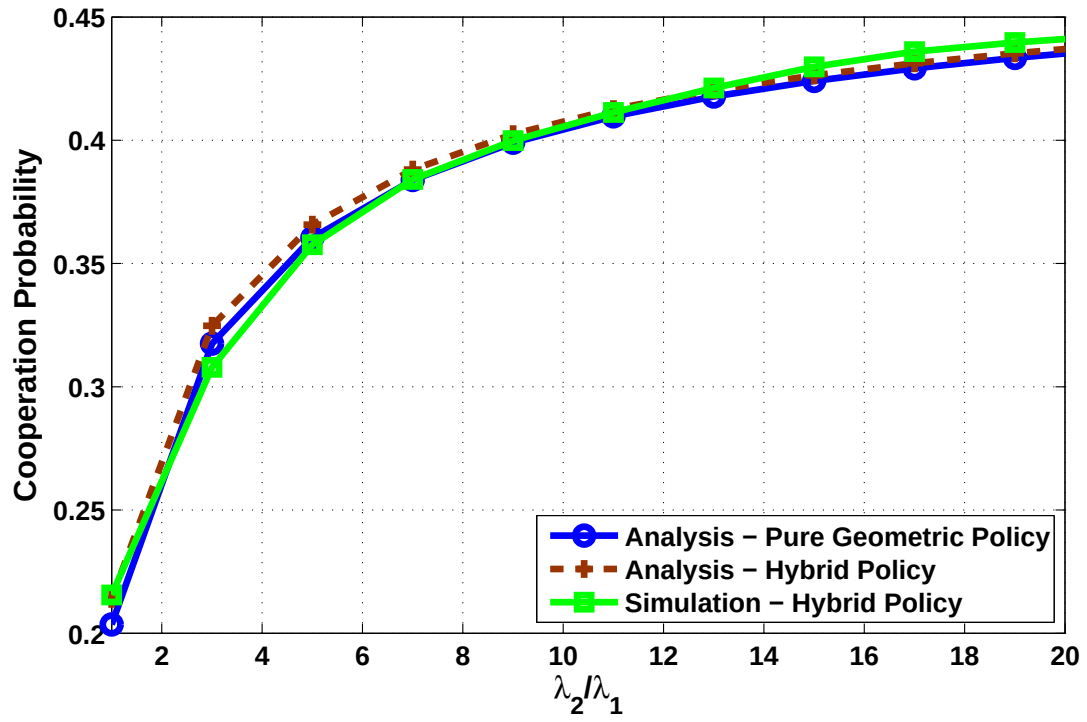

Fig. 6: Cooperation probabilities of different policies versus user density ratio.

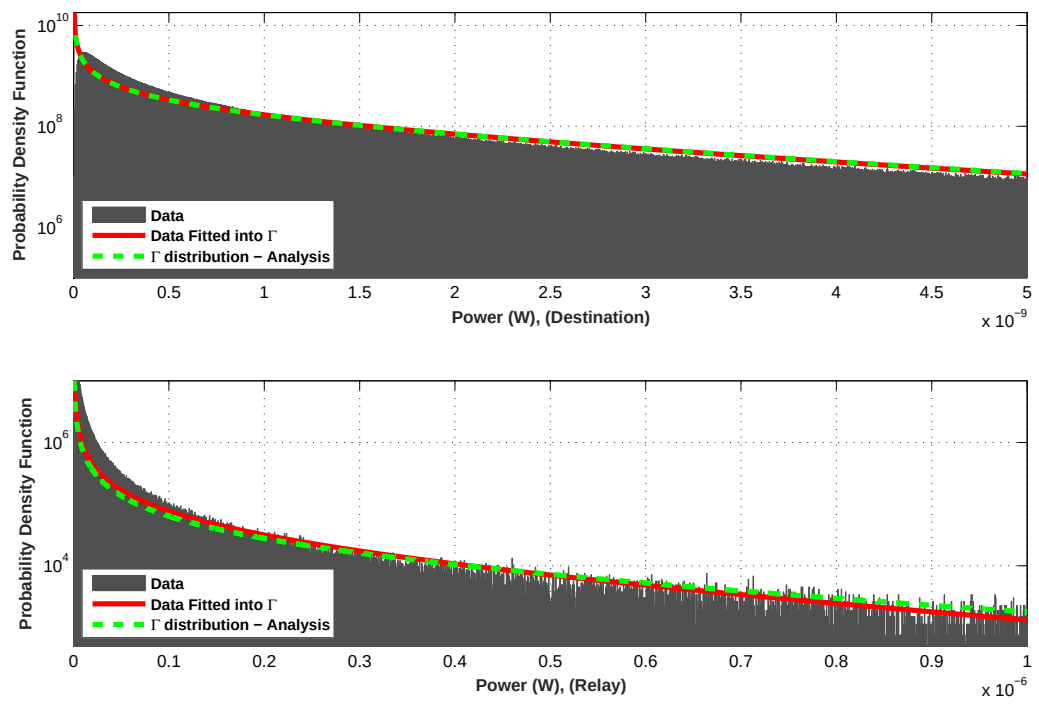

Fig. 7: Gamma distribution fitting of interference power at the destination BS and the relay in the $1^{st}$ phase (SNR = 15 dB).

### *B. Interference Distribution Model Validation*

We now validate the choice of Gamma distribution to model interference power. In Fig. 7, we show two samples of interference power distribution at the destination and at the relay during the first phase. We also show the Gamma distribution fitting to the data numerically obtained through simulation and compare it to the analytically obtained interference power model in Lemma 1 and 2. Both the numerically fitted and the analytic distributions match perfectly. Further, they both can be considered good approximation to the actual interference power distribution for the different network parameters of practical interest as shown and discussed in Figs. 8 and 9.

In Fig. 8, we study the effect of user transmit power on the proposed analytical model for the interference power in two scenarios, when the active user is exactly half way between the BS and the cell edge, and when it is very close to the cell edge, Here we assume that the relaying node is co-located with the active user. The results in Fig. 8 show that the interference power analytical model causes the analytic performance to slightly diverge from simulation only at high transmit power. The maximum transmission power defined in the LTE standard is 23 dBm, and for this range of practical power, the analytic results closely match that of simulation.

In Fig. 9, we study the effect of the locations of both the active user and its associated relay user on the interference power model. Since we assume a fixed-size cell, changing the locations of the active user and the relaying node does not affect the interference model at the destination. Hence, in this study we are mainly concerned with the interference model at the relaying node. Results in Fig. 9 show a close match for all locations of the relay user up to the cell edge. We observe that there is only a single singular point when the relaying node is exactly at the cell edge at which the analytic interference power model fails to capture the actual simulation performance. This event, however, can be practically ignored given the low probability of having an idle node associated with an active user as a relay and located exactly on the cell edge. Fig. 9 further confirms only a slight difference in performance when comparing analytic results to simulation at a transmit power of 26 dBm as also observed in Fig. 8, while the analytic results match simulation perfectly for all other lower transmit power levels.

These validation results suggest that using the Gamma distribution to model the out-of-cell interference power is valid for all range of practical transmit powers and user distances. Next we will use this model to evaluate and analyze the performance gain of network-wide deployment of user-assisted relaying.

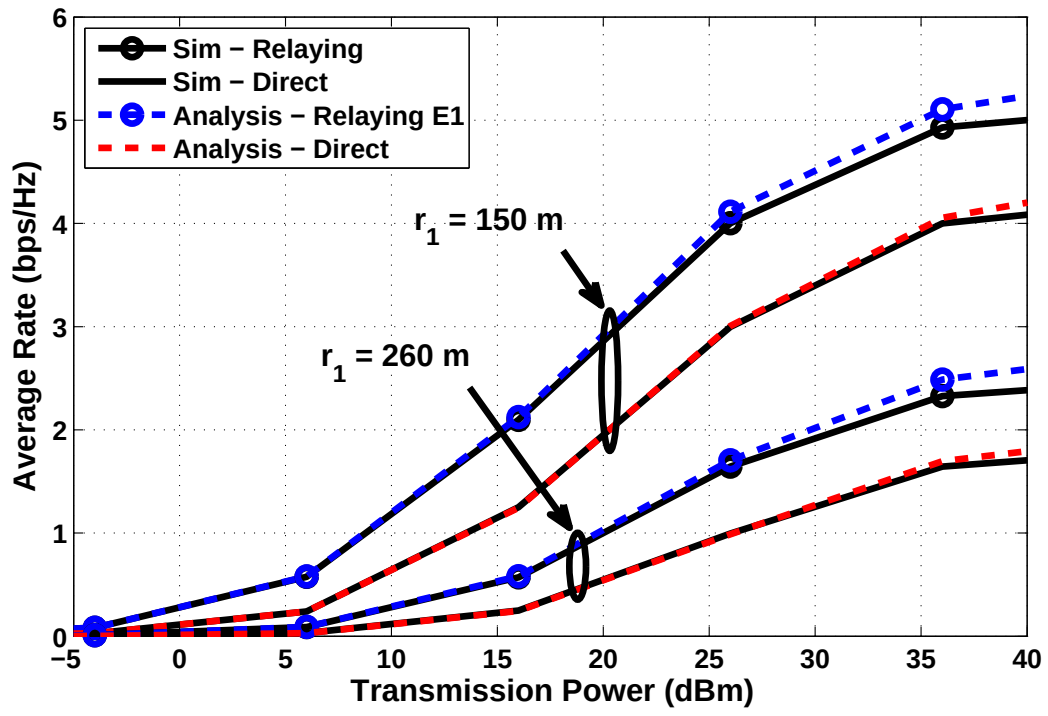

Fig. 8: Gamma interference approximation effect versus transmission power ($r_2 = 0$ m, $R_c = 300$ m).

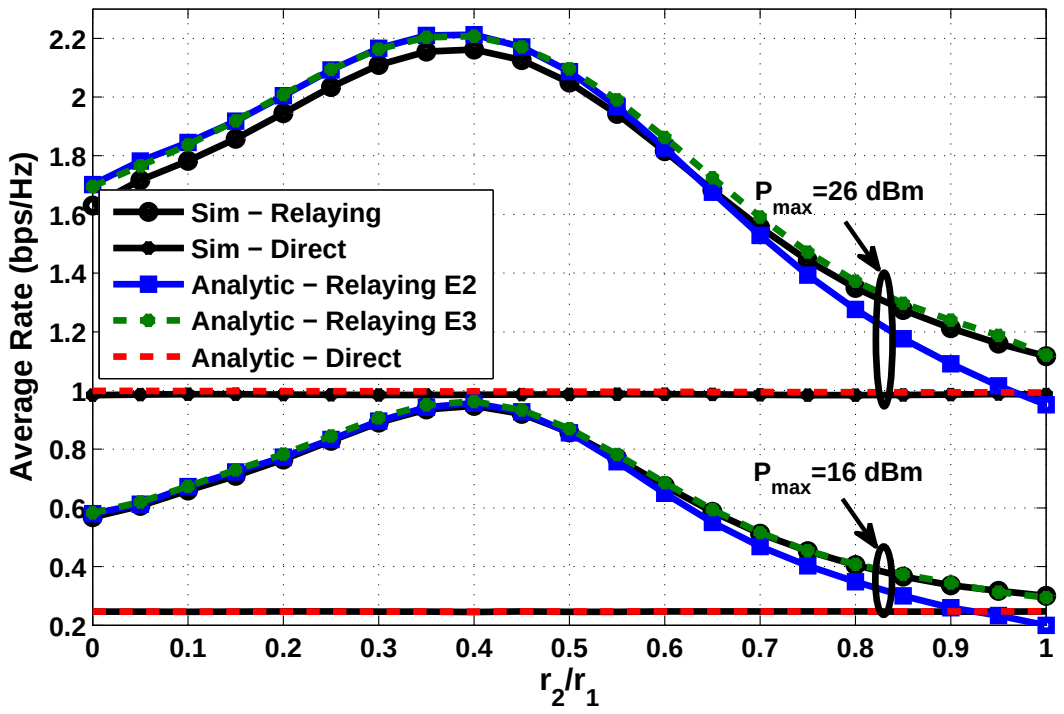

Fig. 10: Average transmission rate versus ratio between cooperation and direct distances for a cell edge user ($r_1 = 260$m, $R_c = 300$m).

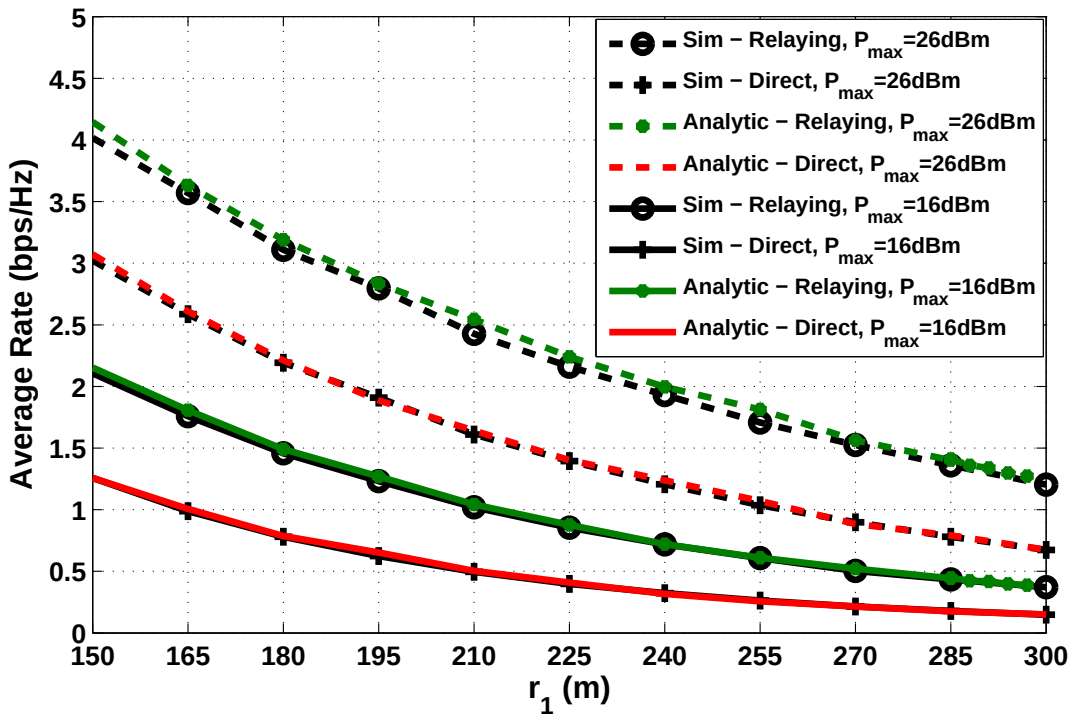

Fig. 9: Gamma interference approximation effect versus direct distance ($r_2 = 0$ m, $R_c = 300$ m).

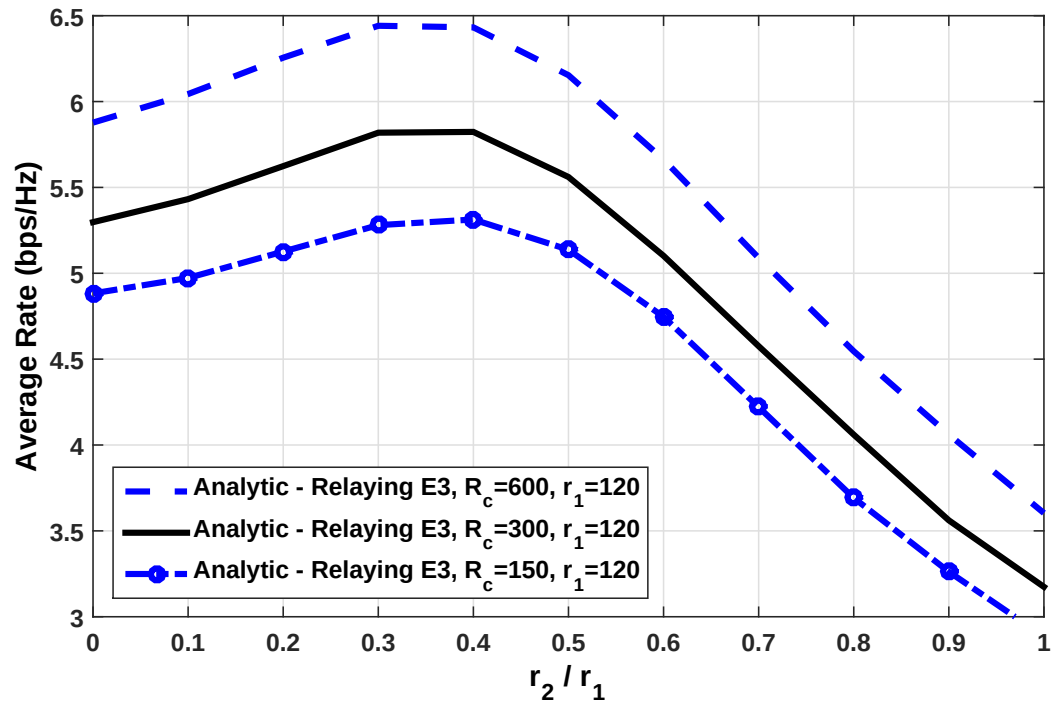

Fig. 11: Average transmission rate versus ratio between cooperation and direct distances for different cell sizes ($r_1 = 120$m, $R_c \in \{150, 300, 600\}$m, $P_{max} = 26$ dBm).

## VII. Performance Analysis of User-Assisted Relaying

In this section, we evaluate the performance of the PDF user-assisted relaying scheme when deploying in the network, taking into account cooperation decisions and out-of-cell interference. We use the transmission average rate as the performance metric. Consider an active UE located within a typical cell of radius $R_c = 300m$, which is proportional to the active user density $\lambda_1 = 1/(16 \times 150^2)$ as discussed earlier in Section V. We present numerical results with $P_{s,i} = P_{r,i}$; $P_{s,i}^b = P_{s,i}^m$; and $P_{s,i}^{m1}$ and $P_{s,i}^{m2}$ allocated optimally to maximize the transmission rate of the active user. Here we perform numerical integrations to compute the average rates in Eqs. (7)−(9) and (16) for the equivalent channel model in Section III-D, based on our developed analytical interference model in Section V-B and cooperation probabilities in Eqs. (25)−(26). We then compare these numerical results with system simulation where we carry out detailed simulation of a multi-cell network as modeled in Section III.

### A. Rate Gain versus Relay Location

In Fig. 10, we compare with simulation results the average rate numerically obtained using our interference power analytic models and the developed cooperation policies versus the ratio between source-to-relay and source-to-destination distances. Note that the cooperation probabilities all approach 0.5 instead of 1 because of the constraining condition $D \leq r_1$ in policies $E_2$ and $E_3$. These results show that performance of the system using the pure geometric cooperation policy, $E_2$, gets worse than simulation using the ideal cooperation policy, $E_1$, when the source-to-relay and source-to-destination distances ratio, $r_2/r_1$, is above 65%, i.e. when the relay is closer to the destination than the source. This difference is mainly due to the lack of the small scale fading information at the transmitting node in deciding whether to perform cooperation or not. An improvement to the performance is observed when we make use of the small scale fading knowledge as in the hybrid cooperation policy, $E_3$, which exhibits a close match with simulation results. Note that there is a slight difference between the analytic and simulation performance results in Fig. 10 at the transmit power 26 dBm, which is mainly due to our exploit of the high transmission power as observed and discussed in Fig. 7. At lower transmit power, this difference almost vanishes as also shown in Fig. 10.

Since cell sizes can vary slightly in practical cellular networks deployment, we examine in Fig. 11 the effect of cell size on performance under policy $E_3$ by changing the radius of the cell under consideration both below and above the typical average value given the density $\lambda_1$. For an active user at a fixed distance from the destination BS, as the cell radius increases, the effective interference at both the relaying node and the destination BS is reduced and hence the performance improves as expected from intuition. This result can also be interpreted in terms of reduced cell size and transmit power. As we decrease the cell size, the amount of transmit power required to reach a user at the cell edge is also decreased. Consequently, the out-of-cell interference is reduced, leading to similar overall performance. In other words, we can reduce both cell size and transmit power without affecting the performance of user-assisted relaying.

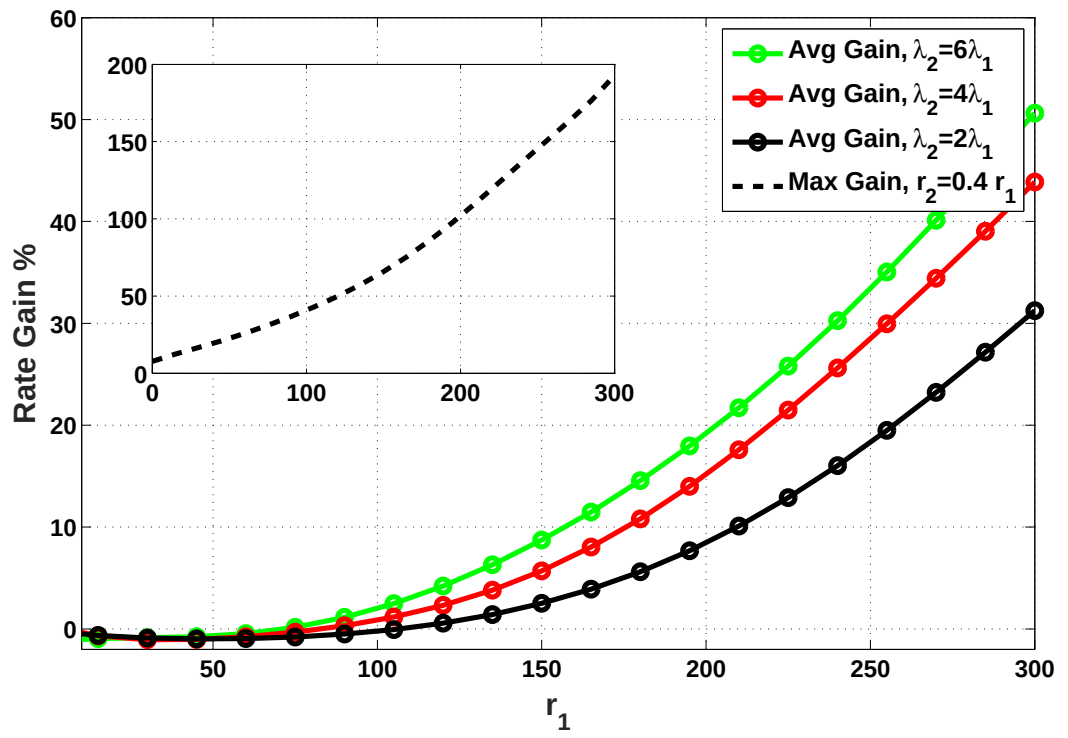

Fig. 12: Average transmission rate gain versus direct distance from the active user to the BS ($R_c = 300$m$, P_{max} = 23$ dBm).

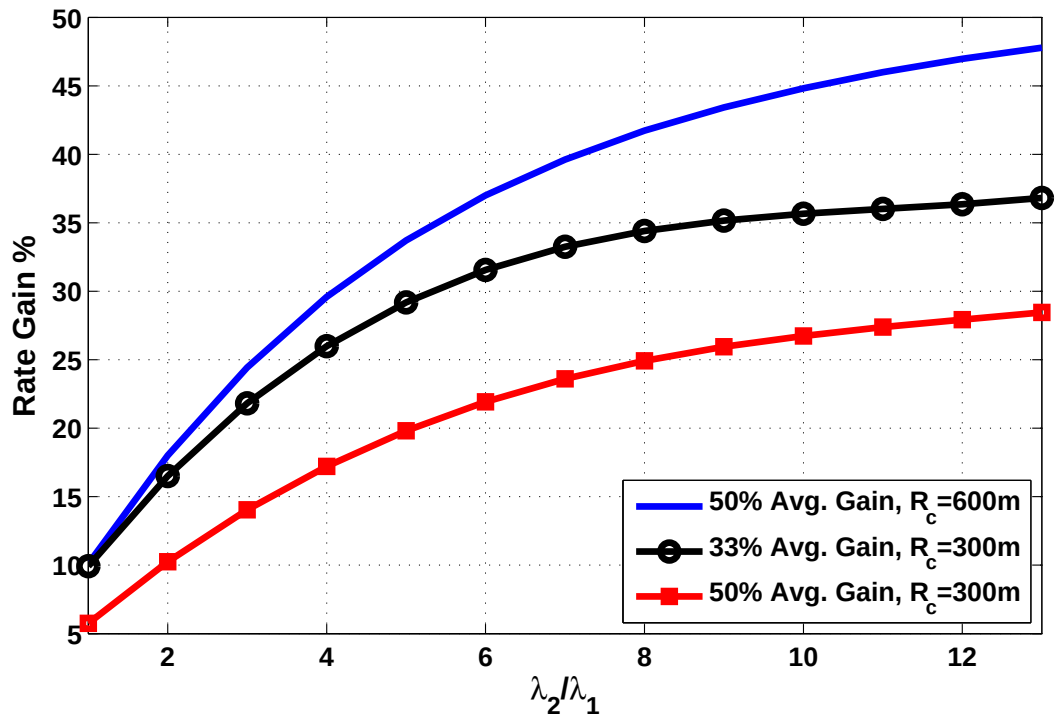

Fig. 13: Average transmission rate gain versus users density ratio ($R_c \in \{300, 600\}$m$, P_{max} = 23$ dBm).

Both Fig. 10 and Fig. 11 show that it is usually more beneficial to have the relaying node closer to the active user than to the destination BS especially when we lack to the knowledge of the small scale fading of the channel. The maximum gain is achieved when the relay user is approximately midway (about 0.4 of the distance) between the active user and the BS.

### *B. Rate Gain Averaged over all User Locations*

Next, we study the performance gain from user-assisted relaying when averaging over the locations of all relay users, and also the locations of all active users. These results are obtained using our analytic results as simulations would require too extensive processing. Further, we have verified the accuracy of the analytic results in Figs. $4-10$.

Fig. 12 shows the rate gain averaged over all possible locations of the relaying user versus the distance from the active user to the destination BS and compare it to maximum gain of the ideal case where the active user always finds a relaying node at exactly $0.4$ distance between the active user and the destination BS, as suggested by results in Figs. 10 and 11. We note that uplink user-assisted relaying poses significant gains for near the cell-edge users while can be irrelevant for active users close to the destination BS. We can also note a slight loss in rate when the active user is very close to the BS, which can be due to the fact that this active user will suffer from the increase out-of-cell interference while not benefiting much from the relaying transmission due to the low probability of finding a relay node at a distance less than the direct distance between the active user and the BS. However, this loss is negligible (less than $1\%$) for users near the center who already have a strong link to the BS, while the whole network throughput gain is significant. This result also suggests user-assisted relaying is most applicable to the farthest $33\%$ or $50\%$ percentile of active users in the cell. For example, at six times more idle users than active users, uplink user-assisted relaying can achieve an average gain of up to $50\%$, and a maximum gain of up to $200\%$ when the idle user is ideally located about halfway between the active user and the base station.

Fig. 13 provides a quantitative evaluation of the average rate gain obtained for active users occupying the farthest $1/3$ and $1/2$ of the cell radius, averaged over all the possible locations of both the active and relay users. The gain increases with higher density of idle users, suggesting that user-assisted relaying is suitable for crowded population areas. These results show that when applying our scheme to the users towards the cell edge, we can achieve higher percentage gain. Also, as the cell radius increases, user-assisted relaying becomes more beneficial to the cell edge users and brings higher rate gains. The average rate gain is almost $30\%$ for the active users located on the one-half of cell radius towards the cell edge, and can increase to almost $50\%$ when the cell size is doubled.

## VIII. Conclusion

In this paper, we analyze system-wide performance impact of deploying user-assisted partial decode-and-forward relaying in a cellular network. Using a stochastic geometry model for user and base station locations, we analytically derive the probability of cooperation and inter-cell interference power generated to the relayed user and destination base station. This cooperation and interference analysis provides a solid theoretical basis for evaluating system performance metrics such as the average transmission rate. Numerical results verify our analysis and show that user-assisted relaying can significantly improve per-user transmission rate, despite increased out-of-cell interference. The transmission rate gain increases with higher idle user density and is more significant for active users closer to the cell edge, suggesting that user-assisted relaying is viable for crowded population areas to improve data rate of near-cell-edge users.

## APPENDIX A: PROOF OF THEOREM 1

In this section, we derive the probability of both the geometric cooperation policy, $E_2$, in (22) and the hybrid fading and geometric cooperation policy, respectively, which can be one of the representations of the cooperation probability $\rho_1$. First, we show the derivation of the policy $E_2$ probability $\rho_2$ detailed as follows

$$\begin{aligned} \rho_2 &= \mathbb{P}\{E_2\} = \mathbb{P}\big\{r_1 \geq r_2, r_1^2 + r_2^2 - 2r_1 r_2 \cos\psi_0 \leq r_1^2\big\} \\ &= \mathbb{P}\big\{r_1 \geq r_2, r_2 \leq 2r_1 \cos\psi_0\big\} \\ &= \int_{-\pi/2}^{-\pi/3} \mathcal{E}_1 d\psi_0 + \int_{\pi/3}^{\pi/2} \mathcal{E}_1 d\psi_0 + \int_{-\pi/3}^{\pi/3} \mathcal{E}_2 d\psi_0, \quad (50) \end{aligned}$$

$$\begin{aligned} \text{where } \mathcal{E}_1 &= 2\pi\lambda_1\lambda_2 \int_0^\infty \int_0^{2r_1\cos\psi_0} r_1 r_2 e^{-\pi(\lambda_1 r_1^2 + \lambda_2 r_2^2)} dr_2 dr_1 \\ &= \frac{2\lambda_2 \cos^2\psi_0}{\pi(\lambda_1 + 4\lambda_2\cos^2\psi_0)}, \quad (51) \\ \mathcal{E}_2 &= 2\pi\lambda_1\lambda_2 \int_0^\infty \int_0^{r_1} r_1 r_2 e^{-\pi(\lambda_1 r_1^2 + \lambda_2 r_2^2)} dr_2 dr_1 \\ &= \frac{\lambda_2}{2\pi(\lambda_1 + \lambda_2)}, \quad (52) \end{aligned}$$

substituting Eqs. (51) and (52) into Eq. (50), we obtain Eq.(25) in Theorem 1.

Now, we derive the probability $\rho_3$ of the hybrid fading and geometric policy $E_3$

$$\begin{aligned} \rho_3 &= \mathbb{P}\{E_3\} \\ &= \mathbb{P}\Big\{r_2 \leq \left(\frac{g_{sr}}{g_{sd}}\right)^{1/\alpha} r_1, r_1^2 + r_2^2 - 2r_1 r_2 \cos\psi_0 \leq r_1^2\Big\} \\ &= \mathbb{P}\big\{r_2 \leq \beta r_1, r_2 \leq 2r_1\cos\psi_0\big\} \\ &= \int_0^2 f_\beta(z) \Bigg[\int_{-\pi/2}^{-\cos^{-1}(z/2)} \mathcal{E}_1 d\psi_0 + \int_{\cos^{-1}(z/2)}^{\pi/2} \mathcal{E}_1 d\psi_0 \\ &+ \int_{-\cos^{-1}(z/2)}^{\cos^{-1}(z/2)} \mathcal{E}_2 d\psi_0\Bigg] dz + \int_2^\infty \int_{-\pi/2}^{\pi/2} \mathcal{E}_1 d\psi_0 dz \\ &= \int_0^2 f_\beta(z) \Bigg[\int_{-\pi/2}^{-\cos^{-1}(z/2)} \mathcal{E}_1 d\psi_0 + \int_{\cos^{-1}(z/2)}^{\pi/2} \mathcal{E}_1 d\psi_0 \\ &+ \frac{\lambda_2 \cos^{-1}(z/2)}{\pi(\lambda_1 + \lambda_2)}\Bigg] dz + \int_2^\infty \int_{-\pi/2}^{\pi/2} \mathcal{E}_1 d\psi_0 dz, \quad (53) \end{aligned}$$

where $\beta = \left(\frac{g_{sr}}{g_{sd}}\right)^{1/\alpha}$ and $f_\beta(z)$ is the probability density function (PDF) of $\beta$.

To obtain the PDF of $\beta$, we first derive the cumulative distribution function (CDF), $F_\beta(z)$, as

$$\begin{aligned} F_\beta(z) &= \mathbb{P}\bigg\{\left(\frac{x_1}{x_2}\right)^{1/\alpha} \leq z\bigg\} = \mathbb{P}\Big\{x_1 \leq z^\alpha x_2\Big\} \\ &= \int_0^\infty \int_0^{z^\alpha x_2} e^{-(x_1+x_2)} dx_1 dx_2 \\ &= \int_0^\infty e^{-x_2}\left(1 - e^{-z^\alpha x_2}\right) dx_2 \\ &= 1 - \frac{1}{1+z^\alpha}, \quad z \in [0,\infty). \quad (54) \end{aligned}$$

The PDF $f_\beta(z)$ is then obtained by differentiating $F_\beta(z)$, as follows

$$f_\beta(z) = \frac{dF_\beta(z)}{dz} = \frac{\alpha z^{\alpha-1}}{(1+z^\alpha)^2}, \quad z \in [0,\infty). \quad (55)$$

## APPENDIX B: PROOF OF THEOREM 2

In this section we derive the *Laplace* transform of the different interference power terms which is used to characterize the moments of interference. The developed moments are then used to develop the interference power distribution analytic model. We first develop the *Laplace* transform of the interference power at the destination during the $1^{st}$ phase as in eq. (56) where the last equality follows from the *Laplace* functional expression for PPP using polar coordinates and assuming the field of interferers outside a cell of fixed radius $R_c$ as discussed in Section V; $\mathcal{L}_G(s)$=$1/(1+s)$ is the Laplace transform of an exponential random variable $G \sim \exp(1)$ and $\mathcal{L}_{\mathcal{J}_{d,i}^b}(s, \|\mathbf{z}_k\|_2)$ is expressed as seen in Eq. (35).

Now, we develop the *Laplace* transform of the interference power at the destination during the $2^{nd}$ phase as in eq. (57) where the last equality also follows from the *Laplace* functional expression for PPP using polar coordinates along with the fixed cell radius assumption; and $\mathcal{L}_{\mathcal{J}_{d,i}^m}(s, \|\mathbf{z}_k\|_2)$ is as in Eq. (36).

Finally, the *Laplace* transform of the interference power at the relay during the $1^{st}$ phase can be developed in a similar way to the case at the destination during the $1^{st}$ phase.

$$\begin{aligned}\mathcal{L}_{\mathcal{Q}^b_{d,i}}(s) &= \mathbb{E}_{\mathcal{Q}^b_{d,i}}\left[e^{-s\mathcal{Q}^b_{d,i}}\right] = \mathbb{E}_{\mathcal{Q}^b_{d,i}}\left[\prod_{\mathbf{z}_k\in\Phi_1\backslash\mathbf{z}_i} e^{-sB_k\left|h^{(k,i)}_{sd}\right|^2 P^b_{s,k} - s(1-B_k)\left|h^{(k,i)}_{sd}\right|^2 P_{s,k}}\right] \\
&= \mathbb{E}_{\Phi_1, g^{(k,i)}_{sd}}\left[\prod_{\mathbf{z}_k\in\Phi_1\backslash\mathbf{z}_i} \mathbb{E}_{B_k} e^{-sB_k g^{(k,i)}_{sd}\|\mathbf{z}_k\|_2^{-\alpha} P^b_{s,k} - s(1-B_k)\|\mathbf{z}_k\|_2^{-\alpha} g^{(k,i)}_{sd} P_{s,k}}\right] \\
&= \mathbb{E}_{\Phi_1, g^{(k,i)}_{sd}}\left[\prod_{\mathbf{z}_k\in\Phi_1\backslash\mathbf{z}_i} \rho_1 e^{-s g^{(k,i)}_{sd}\|\mathbf{z}_k\|_2^{-\alpha} P^b_{s,k}} + (1-\rho_1) e^{-s\|\mathbf{z}_k\|_2^{-\alpha} g^{(k,i)}_{sd} P_{s,k}}\right] \\
&= \mathbb{E}_{\Phi_1}\left[\prod_{\mathbf{z}_k\in\Phi_1\backslash\mathbf{z}_i} \rho_1 \mathcal{L}_G\left(s\|\mathbf{z}_k\|_2^{-\alpha} P^b_{s,k}\right) + (1-\rho_1)\mathcal{L}_G\left(s\|\mathbf{z}_k\|_2^{-\alpha} P_{s,k}\right)\right] \\
&= \mathbb{E}_{\Phi_1}\left[\prod_{\mathbf{z}_k\in\Phi_1\backslash\mathbf{z}_i} \mathcal{L}_{\mathcal{J}^b_{d,i}}\left(s, \|\mathbf{z}_k\|_2\right)\right] = \exp\left(-2\pi\lambda_1\int_{R_c}^{\infty}\left(1-\mathcal{L}_{\mathcal{J}^b_{d,i}}(s,r)\right) r dr\right), \end{aligned} \tag{56}$$

$$\begin{aligned}\mathcal{L}_{\mathcal{Q}^m_{d,i}}(s) &= \mathbb{E}_{\mathcal{Q}^m_{d,i}}\left[e^{-s\mathcal{Q}^m_{d,i}}\right] = \mathbb{E}_{\mathcal{Q}^m_{d,i}}\left[\prod_{\mathbf{z}_k\in\Phi_1\backslash\mathbf{z}_i} e^{-sB_k\left(\left|h^{(k,i)}_{sd}\right|^2 P^m_{s,k} + \left|h^{(k,i)}_{rd}\right|^2 P^m_{r,k}\right) - s(1-B_k)\left|h^{(k,i)}_{sd}\right|^2 P_{s,k}}\right] \\
&= \mathbb{E}_{\Phi_1}\left[\prod_{\mathbf{z}_k\in\Phi_1\backslash\mathbf{z}_i} \mathbb{E}_{g^{(k,i)}_{sd}}\left[\rho_1 e^{-s\|\mathbf{z}_k\|_2^{-\alpha}\left(g^{(k,i)}_{sd} P^m_{s,k} + g^{(k,i)}_{rd} P^m_{r,k}\right)} + (1-\rho_1) e^{-s\|\mathbf{z}_k\|_2^{-\alpha} g^{(k,i)}_{sd} P_{s,k}}\right]\right] \\
&= \mathbb{E}_{\Phi_1}\left[\prod_{\mathbf{z}_k\in\Phi_1\backslash\mathbf{z}_i} \rho_1 \mathcal{L}_G\left(s\|\mathbf{z}_k\|_2^{-\alpha} P^m_{s,k}\right)\mathcal{L}_G\left(s\|\mathbf{z}_k\|_2^{-\alpha} P^m_{r,k}\right) + (1-\rho_1)\mathcal{L}_G\left(s\|\mathbf{z}_k\|_2^{-\alpha} P_{s,k}\right)\right] \\
&= \mathbb{E}_{\Phi_1}\left[\prod_{\mathbf{z}_k\in\Phi_1\backslash\mathbf{z}_i} \mathcal{L}_{\mathcal{J}^m_{d,i}}\left(s, \|\mathbf{z}_k\|_2\right)\right] = \exp\left(-2\pi\lambda_1\int_{R_c}^{\infty}\left(1-\mathcal{L}_{\mathcal{J}^m_{d,i}}(s,r)\right) r dr\right), \end{aligned} \tag{57}$$

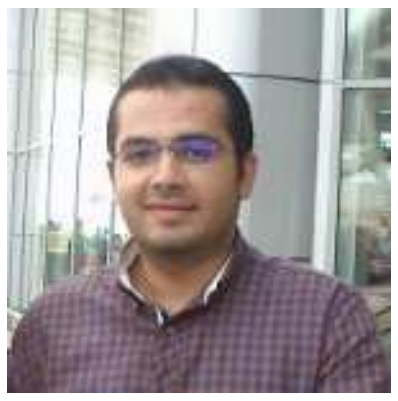

**Hussain Elkotby** is currently pursuing a Ph.D. degree in Electrical Engineering at Tufts University, USA. He received a M.Sc. (2013) and a B.Sc. (2009) degrees in Electrical Engineering from Port Said University, Egypt and Suez Canal University, Egypt, respectively. He finished his undergraduate studies at the top of his class (100 students) and thus was nominated to be a Teaching Assistant in the Department of Electrical Engineering at Port Said University, Egypt. He also worked as a Research Assistant in the Department of Electronics and Electrical Communications Engineering at Cairo University, Egypt and contributed to the 4G++ research project. Hussain's research interests are in the general area of cellular networks and wireless communications with current focuses on stochastic geometry, cooperative communications, device-to-device communications, and interference alignment.

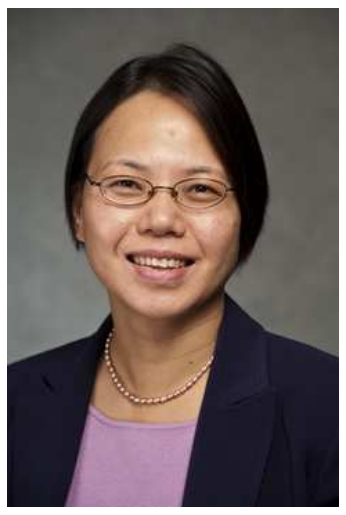

**Mai Vu** (M'06-SM'13) received the Ph.D. degree in electrical engineering from Stanford University, USA, after receiving the M.S.E. degree in electrical engineering from the University of Melbourne, Australia, and a bachelors degree in computer systems engineering from the Royal Melbourne Institute of Technology (RMIT University), Australia. Between 2006 and 2008, she worked as a Lecturer and Researcher at the School of Engineering and Applied Sciences, Harvard University. During 2009-2012, she was an Assistant Professor in Electrical and Computer Engineering at McGill University. Since January 2013, she has been an Associate Professor in the Department of Electrical and Computer Engineering at Tufts University.

Dr. Vu conducts research in wireless systems, signal processing, and networked communications. She has published extensively in the areas of cooperative and cognitive communications, relay networks, MIMO capacity and precoding, and energy-efficient communications. She has served on the technical program committee of numerous IEEE conferences and is currently an editor for the IEEE TRANSACTIONS ON WIRELESS COMMUNICATIONS